\documentclass[twocolumn]{aa}  
\usepackage{graphicx}
\usepackage{color}
\usepackage{epsfig}
\usepackage{supertabular}
\usepackage{rotating}
\usepackage{natbib}
\usepackage{txfonts}
\begin{document}
   \title{Two formation channels of UCDs in Hickson Compact Groups \thanks{Based on observations obtained in service mode at the VLT (programme 082.B-0882)}}

   %\subtitle{UCDs in HCGs}

   \author{
          C. Da Rocha
          \inst{1,2}
          \and
          S. Mieske
          \inst{3}
          \and
          I. Y. Georgiev 
          \inst{4}
          \and
          M. Hilker 
          \inst{2}
          \and
          B. L. Ziegler
          \inst{2}
          \and
          C. Mendes de Oliveira  
          \inst{5}
         }

   \offprints{C. Da Rocha}

   \institute{
              N\'ucleo de Astrof\'{\i}sica Te\'orica, Universidade Cruzeiro 
              do Sul, R. Galv\~ao Bueno 868, 01506--000, S\~ao Paulo, SP, Brazil
         \and
              European-Southern Observatory, Karl-Schwarzschild Str. 2, 85748 
              Garching, Germany
         \and
              European Southern Observatory, Alonso de Cordova 3107, 
              Vitacura, Santiago, Chile
         \and
              Argelander Institut f\"ur Astronomie, Universit\"at Bonn, Auf dem
              H\"ugel 71, 53121 Bonn, Germany
         \and
              Instituto de Astronomia, Geof\'{\i}sica e Ci\^encias
              Atmosf\'ericas, Universidade de S\~ao Paulo, Rua do Mat\~ao 1226,\\
              Cidade Universit\'aria, 05508--900, S\~ao Paulo, SP, Brazil
         }

   \date{}

% \abstract{}{}{}{}{} 
% 5 {} token are mandatory
 
  \abstract % context heading (optional) 
{The formation of ultra-compact dwarf galaxies (UCDs) is believed
to be interaction driven, and UCDs are abundant in the cores of
galaxy clusters, environments that mark the end-point of galaxy
evolution. Nothing is known about the properties of UCDs in compact
groups of galaxies, environments where most of galaxy evolution and 
interaction is believed to occur and where UCDs in intermediate state 
of their evolution may be expected.}
%leave
%it empty if necessary
% aims heading  (mandatory) 
{The main goal of this study is to detect and characterize, for the
first time, the UCD population of compact groups of galaxies.  For
that, two nearby groups in different evolutionary stages, HCG\,22
and HCG\,90, were targeted.}
% methods heading  (mandatory) 
{We selected about 40 UCD candidates from pre-existing photometry
of both groups, and obtained spectra of these candidates using the
VLT FORS2 instrument in MXU mode. Archival HST/ACS imaging was used
to measure their structural parameters.}
% results heading (mandatory) 
{We detect 16 and 5 objects belonging to HCG\,22 and HCG\,90,
respectively, covering the magnitude range $-10.0>M_R>-11.5$ mag.
Their integrated colours are consistent with old ages covering a
broad range in metallicities (metallicities confirmed by the
spectroscopic measurements).  Photometric mass estimates put 4 objects
in HCG\,90 and 9 in HCG\,22 in the mass range of UCDs
($>2\times10^6 M_{\odot}$) for an assumed age of 12\,Gyr. These UCDs
are on average 2-3 times larger than the typical size of Galactic
GCs, covering a range of $2\lesssim r_{h}\lesssim 21~\rm pc$.  The
UCDs in HCG\,22 are more concentrated around the central galaxy
than in HCG\,90, at the 99\% confidence level. They cover a broad
range in [$\alpha$/Fe] abundances from sub- to super-solar. The
spectra of 3 UCDs (2 in HCG\,22, 1 in HCG\,90) show tentative
evidence for intermediate age stellar populations. The clearest
example is the largest and most massive UCD ($\sim10^7 M_{\odot}$)
in our sample, detected in HCG\,22. Its properties are most consistent
with a stripped dwarf galaxy nucleus. We calculate the specific
frequency ($S_N$) of UCDs for both groups, finding that HCG\,22 has
about three times higher $S_N$ than HCG\,90. }
% conclusions heading (optional),  leave it empty if necessary 
{The ensemble properties of the detected UCDs supports two co-existing
formation channels: a star cluster origin (low-luminosity, compact
sizes, old ages, super-solar $\alpha/$Fe), and an origin as tidally
stripped dwarf nuclei (more extended and younger stellar populations).
Our results imply that the UCDs detected in both groups do not, in
their majority, originate from relatively recent galaxy interactions.
Most of the detected UCDs have likely been brought into the group
together with their host galaxies. }

\titlerunning{UCDs in Hickson Compact Groups}

   \keywords{galaxies: groups: individual: HCG\,22 and HCG\,90 -- galaxies:
dwarf -- galaxies: interactions -- galaxies: star clusters}

   \maketitle 
%
%________________________________________________________________

\section{Introduction}
  Within the last decade, a new class of compact stellar systems,
  called ``ultra-compact dwarf galaxies'' (UCDs), was discovered in
  the cores of the Fornax, Virgo, Centaurus and Hydra galaxy clusters
  \citep{hil99,dri00,mie04a,has05,jon06,evs07a,hil07,mie07b,mie08c,mis08}.
  UCDs are characterized by evolved stellar populations, half-light
  radii of $10 \lesssim r_h \lesssim 100~\rm pc$ and masses of $2\times
  10^6 < m < 10^8 M_{\odot}$ \citep{mie08c}. They are
  hence of intermediate nature between globular clusters (GCs) and
  dwarf elliptical galaxies, covering a mass range starting at about
  that of $\omega$Cen up to almost M32. The dynamical M/L ratios
  of UCDs are on average about twice as large as those of Galactic
  globular clusters at the same metallicity
  \citep[e.g.][]{dab08,mie08e,for08,kru09,tay10}.  This indicates
  that UCDs may mark the on-set of dark matter domination in small
  stellar systems \citep{gil07,goe08}, or probe a variation of the
  IMF \citep{dab08,mie08e}.

  Several formation channels of UCDs have been discussed in the
  literature, of which the first two are directly related to galaxy
  interactions: 1. UCDs have formed from the amalgamation of many
  young massive star clusters in the tidal arms of gas-rich galaxy
  mergers \citep{feh02}; 2. UCDs are remnant nuclei of dwarf galaxies
  which have lost their stellar halo due to interaction with the
  tidal field of the galaxy cluster \citep{bek03}; 3.  UCDs were
  formed together with the main body of their host galaxies' globular
  cluster system, representing the very bright tail of the globular
  cluster luminosity function \citep{mie04a,gre09}.  4. UCDs originate
  from primordial small scale overdensities, and they survived the
  galaxy cluster formation and evolution processes until the present
  time \citep{dri04}.

\vspace{0.2cm}

  Up to now, UCDs have been mainly observed in the cores of evolved
  clusters, like Fornax, Virgo or Centaurus. These environments --
  being dominated by early-type galaxies -- mark the end-point of
  galaxy evolution with relatively large velocity dispersions and
  low galaxy merger rates at present.  A possible very young UCD
  progenitor has been observed in the merger remnant NGC\,7252, and
  UCDs have also been detected in small loose groups, like Dorado,
  CenA and Sombrero \citep{evs07b,hau09,tay10}.

  Surprisingly, no searches for UCDs have been performed in such
  environments where galaxy interactions and transformation occur
  very often, namely in Compact Groups of Galaxies. It is therefore
  the proper environment to test galaxy interaction driven UCD
  formation scenarios. Observations and simulations are showing
  that great part of the galaxy evolution and ``pre-processing''
  happens in groups \citep{fuj04,rud06}.  Merging and subsequent
  infall of such groups leads to the formation of the present-day
  galaxy clusters.  Compact groups are an environment where the
  evolution of galaxies occurs under the most extreme conditions,
  due to their high projected densities and low velocity dispersions
  \citep{hic92}, which implies that galaxy interactions should be
  frequent and efficient in driving the morphological transformation
  of galaxies.  Signs of interaction, like morphologically distorted
  galaxies, are frequently found in compact groups \citep[e.g.][]{men94},
  as well as star formation regions and tidal dwarf galaxy candidates
  \citep{igl01,men01}, an intragroup diffuse light (IGL) component
  \citep{dar05,dar08} all of which shows the effects of galaxy
  interactions in such environments.  Since the formation of UCDs
  is believed to be interaction driven (see above), compact groups
  are an ideal place to look for UCDs and test these proposed
  formation scenarios.  Galaxies in HCGs have globular cluster
  systems (GCSs) typical for the host galaxy luminosity
  \citep{dar02,dar03}.

\vspace{0.2cm}

In this paper we present for the first time a spectroscopic search
for UCDs in the two compact groups HCG\,22 and HCG\,90
\citep[$(m-M)=32.6~\rm mag$ for both groups, corresponding to
$d=33.1~\rm Mpc$,][]{bla01} using the FORS2 instrument at the VLT
in MXU mode, complemented with high resolution imaging from archival
Hubble Space Telescope (HST) WFPC2 and ACS data.

HCG\,90, as originally catalogued by \citet{hic82}, consists of
four giant galaxies, the Sa NGC\,7172 (HCG\,90a), the E0's NGC\,7176
and 7173 (HCG\,90b and c, respectively) and the irregular NGC\,7174
(HCG\,90d), with magnitudes in the range $M_B=-20.3$ and $-19.8$
\citep{hic89}. Twelve smaller member galaxies, fainter than Hickson's
magnitude range, were latter identified by \citet{dec97} and
\citet{zab98a}, leading to a group mean radial velocity of
$2545\pm58~\rm km~s^{-1}$ and velocity dispersion of $193_{-33}^{+36}~\rm
km~s^{-1}$ \citep{zab98a}. The group's angular diameter is about
$9'$.  NGC\,7172 falls out of our observed field of view. The other
three galaxies show optical signs of interaction \citep{men94}, an
ionized gas disk at NGC\,7172, a common warm gas envelope at the
pair NGC\,7176 and 7174 \citep{pla98,cas04} and a diffuse X-ray
halo \citep{mul98,whi03,osm04}.  It has also a significant component
of intragroup diffuse light, constituting about 35\% of its stellar
component of the observed part of the group (Da Rocha \& Ziegler
2010 - in preparation), with colours consistent with old stellar
population, showing that this group is in an advanced evolutionary
stage, which is in agreement with other dynamical evolution indicators
such as its crossing time ($\simeq0.02\ t_{\rm Hubble}$) and low
fraction of spiral galaxies \citep{hic92}.

HCG\,22, as originally catalogued by \citet{hic82}, consists of
three giant galaxies, E2 NGC\,1199 (HCG\,22a) the Sa NGC\,1190
(HCG\,22b) and the SBcd NGC\,1189 (HCG\,22c), with magnitudes in
the range from $M_B=-20.4$ to $-18.1$ \citep{hic89}.  An additional
smaller member galaxy was latter identified by \citet{dec97}, leading
to a group mean radial velocity of $2629\pm33~\rm km~s^{-1}$ and
velocity dispersion of $40\pm28~\rm km~s^{-1}$ \citep{rib98}. The
group's angular diameter is about $6'$. This group is in an earlier
evolutionary stage, as indicated by the low number of optical signs
of galaxy interaction \citep{men94}, its larger crossing time
($\simeq0.2\ t_{\rm Hubble}$), high spiral fraction \citep{hic92}
and H{\sc i} content \citep{ver01}.  This group shows no signs of
intragroup light, where most of the optical luminosity is concentrated
on its brightest galaxy (NGC\,1199). Also the the X-ray emission
in the group is centered on this galaxy \citep{pon96,whi00}.
NGC\,1199, considered to be the dominant galaxy at the group center
is also the galaxy with the richest GCS among the group members.

The differences between the two groups, especially the distribution
of the optical and X-ray emission (and mass, assuming light traces
the mass), and the signs of recent or old interaction are an important
factor in understanding the formation mechanisms of objects like
the UCDs.

\section{Observations and data reduction}

\subsection{Observations}

\begin{figure*}
\begin{center}
  \includegraphics[width=8.6cm]{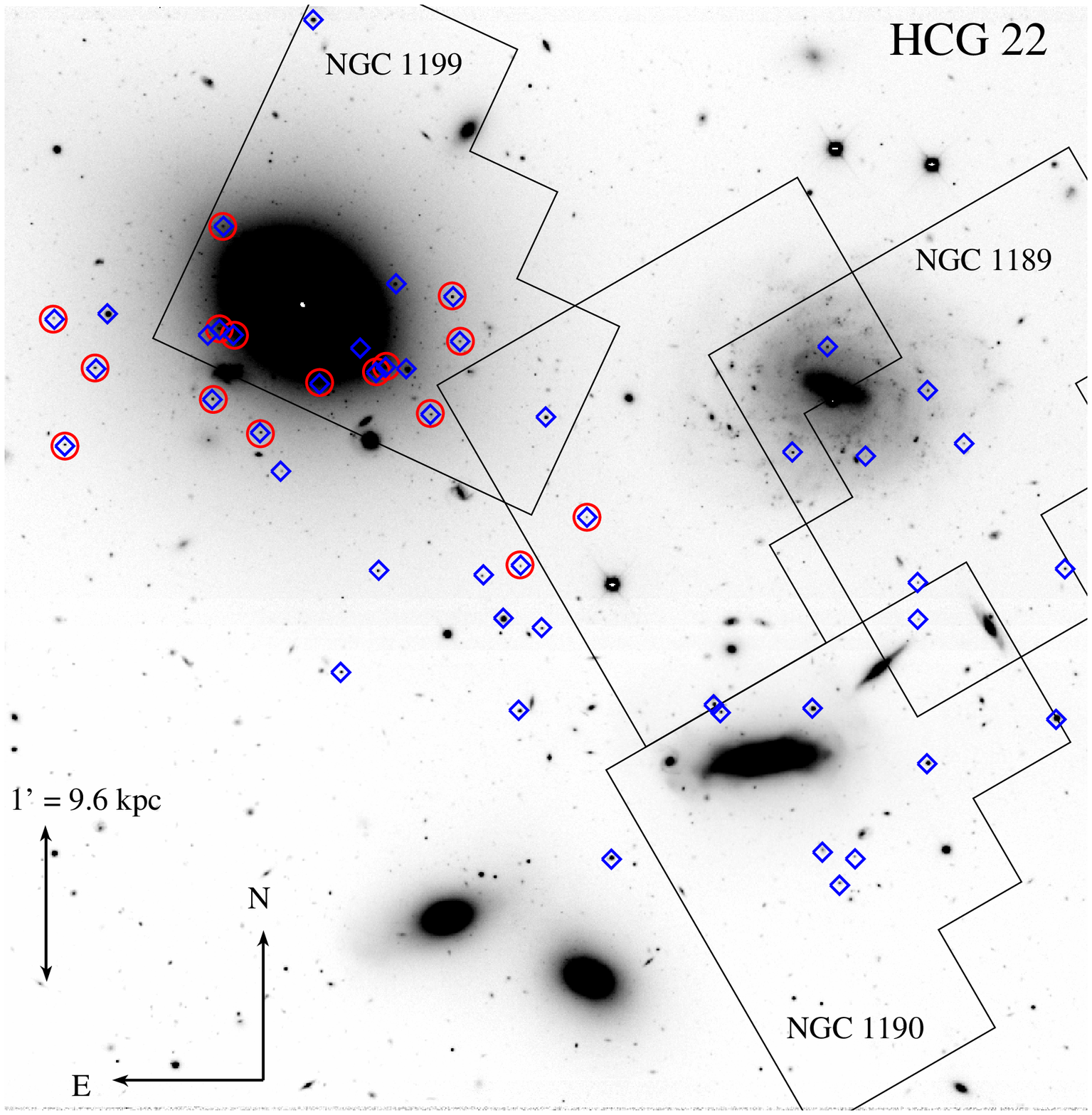}
\hspace{0.5cm}
  \includegraphics[width=8.6cm]{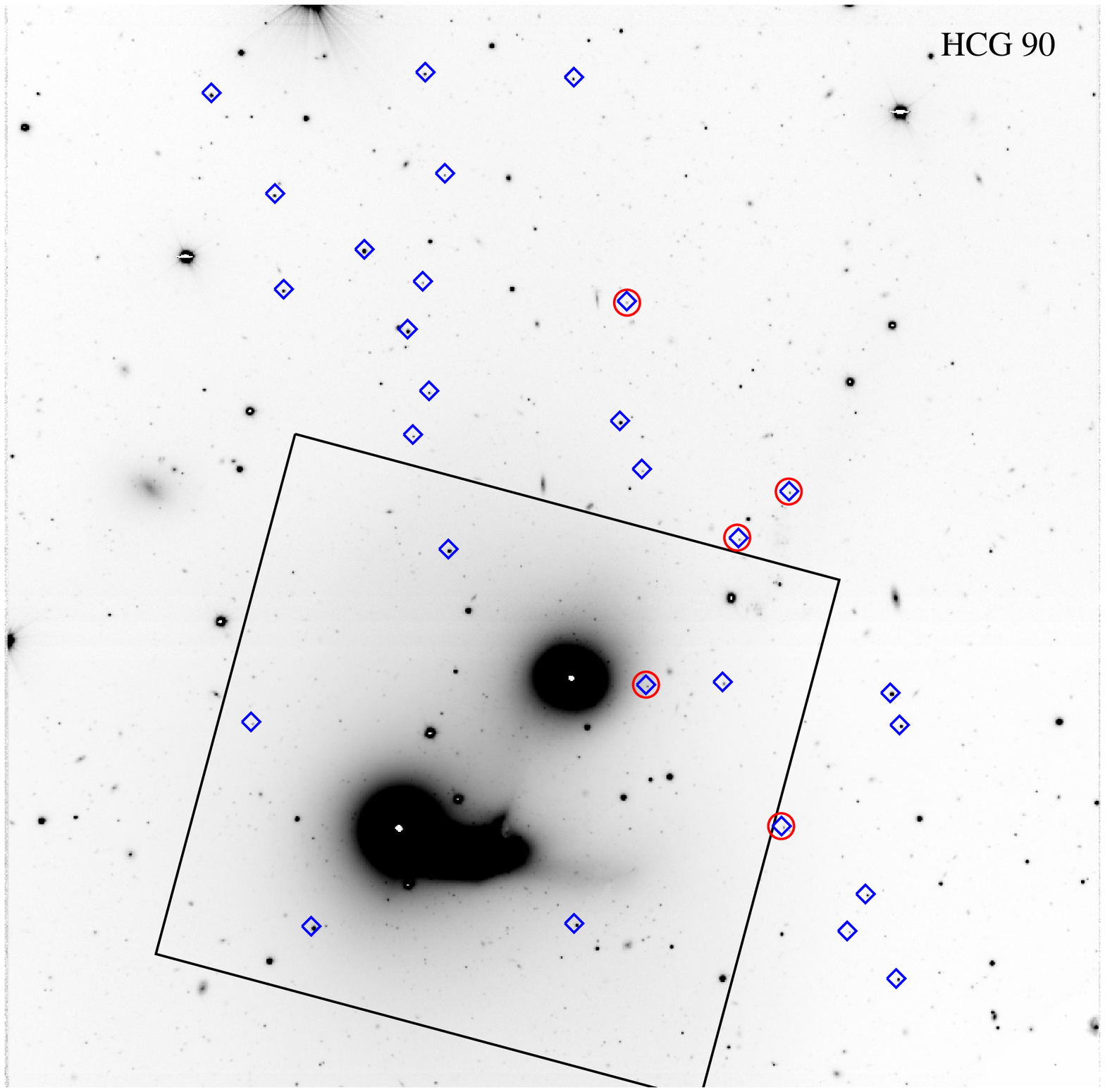}
  \caption{FORS2 $R$-band pre-images of HCG\,22 (left) and HCG\,90
  (right), of dimension $7\times7'$, corresponding to about
  $65\times65~\rm kpc$ at the distance of the groups. Blue diamonds
  show the spectroscopic targets and those marked with red circles
  are confirmed compact members of the HCGs (most in the mass range
  of UCDs), i.e. those with radial velocities within the group range
  (cf.  Figure~\ref{vradhist}). Note the difference in their spatial
  distribution: UCD candidates in HCG\,22 are clustered around the
  central galaxy, while the few UCDs in HCG\,90 have more dispersed
  distribution.  Overlaid are also the HST pointings used in our
  analysis (WFPC2 for HCG\,22 and ACS/WFC for HCG\,90).  Labels in
  HCG\,22 indicate the name of the target at which the WF3 chip was
  centered. Physical scale and orientation are indicated in the lower
  left corner and are the same for both fields. A colour version
  of this figure, showing a high-resolution HST zoom at the dusty
  disk in the central $5~\rm kpc$ of NGC\,1199 and gas dominated
  spiral in HCG\,90, is available as an online material.}
\label{images}
\end{center}
\end{figure*}

The spectroscopic data for this study were obtained in October/November
2008 in MXU mode with the FOcal Reducer and Spectrograph FORS2
\citep{app98} mounted on UT1 Antu at the VLT (programme 082.A-0882,
PI: Da Rocha), using the 600B grism. FORS2 is equipped with two
2k\,x\,4k MIT CCDs (15$\mu$m pixel size), and has a pixel scale of
$0\farcs25$ for the commonly used SR collimator position, providing
a $7\times7'$ field-of-view.  The 600B grism is centered on 465 nm
and provides a resolution of 0.75 {\AA} per pixel. We adopted a
slit-width of $1''$, such that our instrumental resolution was
$\simeq$3{\AA}, corresponding to R$\simeq$1500 or about $200~\rm
km~s^{-1}$ (in terms of FWHM).  For each mask, three exposures of
2914 seconds were taken, with a total of 8742 seconds per mask
(about 2.4 hours).  Figure~\ref{images} shows the $7\times7'$ FORS2
pre-images in the $R$-band of the two target Hickson Compact Groups
22 and 90, each corresponding to a physical scale of about
$65\times65~\rm kpc$ at the adopted distance to HCG\,22 and HCG\,90
of $33.1~\rm Mpc$.

In order to measure the structural properties of our targets we
have accessed HST/ACS and WFPC2 archival imaging data of the two
compact groups.  These images were acquired in 2006 and 2007 as
part of programs GO\,10554 (PI:\,R.\,Sharples) and GO\,10787
(PI:\,J.\,Charlton) for HCG\,90 and HCG\,22, respectively. The
ACS/WFC observations of HCG\,90 consist of one pointing centered
between the two most luminous galaxies (NGC\,7173 and NGC\,7176)
at RA\,$=22^h02^m3^s94$ and DEC\,$=-31^o59'04''34$.  Four dither
line exposures with total integration time of 1375 and 3075 seconds
were taken in F475W ($1\times340, 3\times345$\,seconds) and F850LP
($1\times639,774$ and $2\times831$\,seconds), respectively.  The
HCG\,22 field was covered with four sub-pixel dithered exposures
with WFPC2 totaling 1900 seconds ($1\times400$ and $ 3\times500$\,sec)
in the F450W, F606W and F814W filter set.  The positions and
orientation of the HST observations are overlaid on the FORS2
pre-images in Figure~\ref{images} as well. Both, the ACS and the
WFPC2 imaging programs target the star cluster populations around
the galaxies of the corresponding groups. The ACS imaging exposure
times were designed to reach about a magnitude beyond the GC
luminosity function turnover ($M_{V,\rm TO}\simeq-7.4~\rm mag$)
while the WFPC2 observations reach about two magnitudes above
($M_V\gtrsim-9.0~\rm mag$) at signal-to-noise of 30 and 10,
respectively.  This makes them ideal in depth and resolution for
the purposes of the current study to measure structural and photometric
properties of spectroscopically confirmed UCD candidates.

\subsection{Candidate selection}
\label{sel}
\begin{figure*}
\begin{center}
\includegraphics[width=8.6cm]{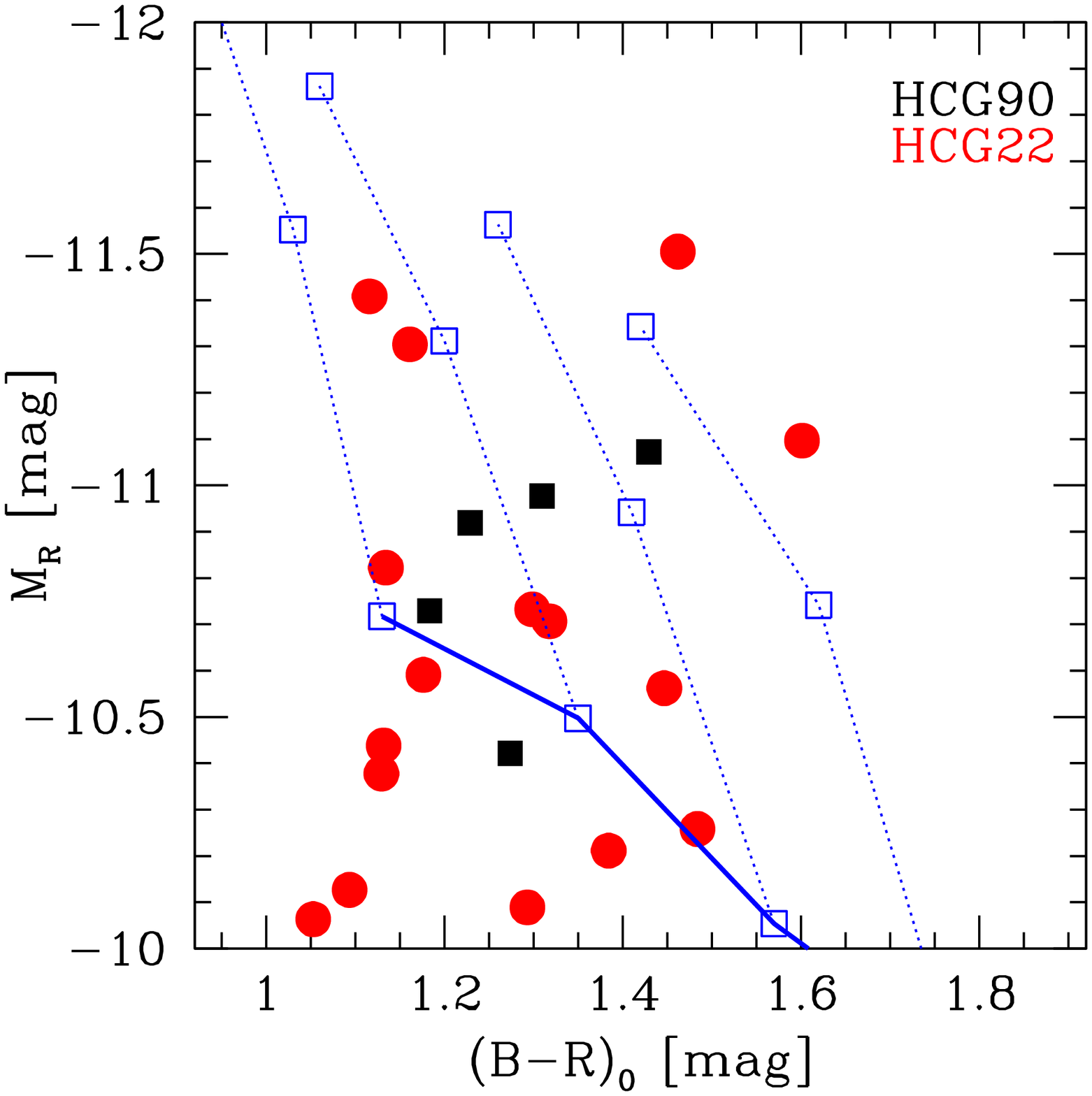}
\includegraphics[width=8.6cm]{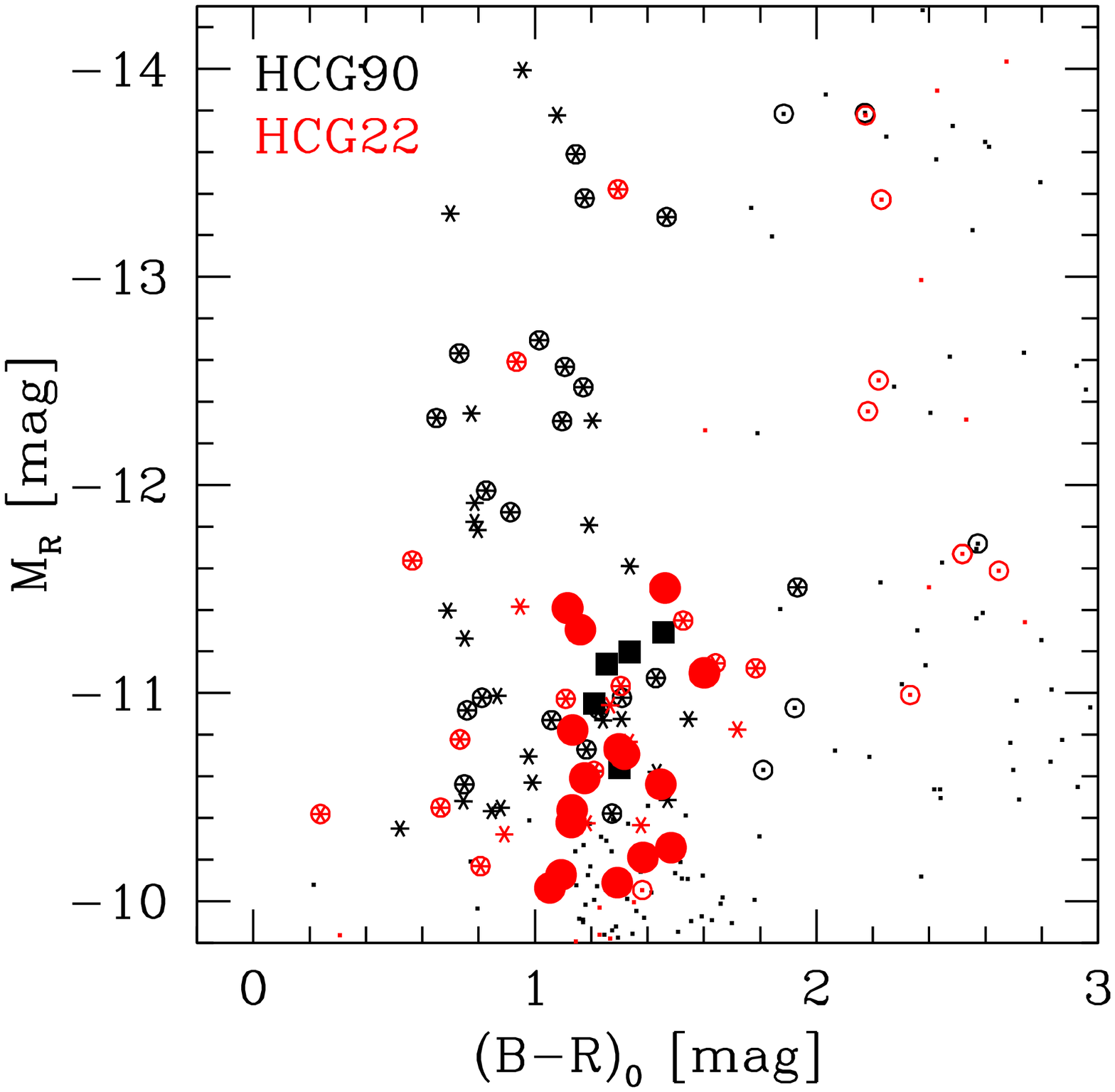}
  \caption{Colour-magnitude diagrams of HCG compact members (left
  panel) and all preselected targets (right panel).  {\bf Left
  panel:} $(B-R)_0$ is plotted vs. $M_R$ for spectroscopically
  confirmed members in the two groups. The dotted lines are model
  predictions for simple stellar populations from \citet{bru03} for
  a range of metallicities and ages, and a stellar mass of $2\times10^6
  M_{\odot}$, which marks the limit between normal GCs and UCDs
  \citep[e.g.][]{mie08c}. From left to right the lines correspond
  to metallicities [Z/H]=-1.6, -0.6, 0.1, and 0.4 dex.  The open
  squares indicate ages of 2, 5, and 12 Gyrs, from top to bottom.
  The solid (blue) line indicates the 12\,Gyr isochrone for a stellar
  mass of $2\times10^6 M_{\odot}$.  {\bf Right panel:} Analogous
  to the left panel, but for a wider colour-magnitude range to show
  the confirmed group members and the preselected targets of our
  survey. Filled red circles and black squares indicate the confirmed
  group members, the same as in the left panel. Open circles indicate
  all other objects that were included in the MXU masks. Sources
  marked by asterisks are those which were high priority candidates,
  according to their colour-magnitude values and morphology. Small
  dots indicate target sources of lower priority (colours fall
  outside the selection color range, or of more resolved morphology).}
\label{CMD}
\end{center}
\end{figure*}

  For both HCGs, we used deep multi-band optical photometry covering
  the central $7'$ ($65~\rm kpc$), in addition to the shallower
  $R$-band pre-imaging shown in Figure~\ref{images}.  For HCG\,22,
  these are $B$ and $R$ images taken with Keck at $\simeq0\farcs8$
  seeing \citep[][ see their Figure~1]{dar02}.  For HCG\,90, these
  are FORS1 archive images of one field in $B$, $V$ and $R$-band
  (programme 65.N-0547(A); Da Rocha \& Ziegler 2010 - in preparation).
  We selected unresolved sources as UCD candidates from these deep
  imaging, and re-observed the same fields in the $R$-band during
  the pre-imaging for our MXU run to obtain accurate relative
  coordinates.  We used the SExtractor star-classifier to separate
  unresolved from resolved sources, and furthermore performed visual
  inspection to cross-check the SExtractor classifications. This
  restriction to unresolved sources corresponds to an upper limit
  of projected half-light radius of about $r_h \simeq~50-60~\rm pc$
  at the distance of HCG\,22 and HCG\,90, encompassing the size
  range of all but two of the known UCD population. In the mask
  creation process we assigned low priorities both to objects too
  red to be at redshift 0 ($(B-R)_0>2.0~\rm mag$) and very blue
  objects ($(B-R)_0<0.6~\rm mag$) to exclude foreground stars. The
  blue limit corresponds to a single stellar population of 1 Gyr
  age at metallicity ${\rm [Fe/H]}=-2.0~\rm dex$, or 0.5 Gyr age
  at metallicity ${\rm [Fe/H]}=0.0~\rm dex$, according to \citet{wor94}
  and \citet{bru03}, thus potential UCD candidates younger than
  0.5-1 Gyr were not included in our spectroscopic sample.  We
  further restrict to objects in the UCD magnitude regime $-14.0 <
  M_{R,0} < -10.0~\rm mag$ ($19.0<R<23.0~\rm mag$) (Figure~\ref{CMD}).
  The faint magnitude limit overlaps with the regime of massive
  globular clusters.

For each group we designed two MXU masks containing most of the
high priority candidates (Figure~\ref{CMD}) For each mask the planned
total exposure time was 2.4 hours.

For HCG\,22, both of these masks were observed in service mode,
yielding spectra for 54 objects, of which seven were in common
between both masks for consistency check. Of the spectra for the
47 individual sources, 10 were too faint for reliable radial velocity
measurement, yielding 37 successfully observed sources of which 29
were of high priority (i.e. fulfilling the colour-magnitude selection
criteria as outlined above). This compared to a total number of 38
high priority sources detected on the pre-images leads to a
completeness of 76\% (29/38) of the high priority candidates in
HCG\,22 with radial velocity measurements.

Only one mask was observed for HCG\,90, containing 28 objects, of
which all got reliable radial velocity measurement. 23 of those
were high priority candidates. This compared to a total number of
47 high priority sources detected on the pre-images, makes our
completeness for HCG\,90 high priority candidates with measured
radial velocities of 49\% (23/47).

\subsection{Data reduction: VLT Spectroscopy}

The data reduction of the MXU spectra was executed with standard
IRAF routines in the TWODSPEC and ONEDSPEC packages. We performed
cosmic ray rejection on each of the three raw 2D spectra for each
mask, using the LACOSMIC algorithm \citep{dok01}. We then averaged
the three single, cosmic ray-cleaned exposures without applying any
further rejection algorithm. Spectrum offsets perpendicular to the
dispersion direction were negligible between the three single
exposures.  The spectra on the combined image were extracted using
the {\sc apall} task in the TWODSPEC package, and wavelength
calibrated using the identify and {\sc dispcor} task in the ONEDSPEC
package.

To measure the radial velocities of our targets, we performed Fourier
cross-correlation with a template spectrum using the IRAF task {\sc
fxcor} in the RV package. As reference template we used a synthetic
spectrum of the stellar population of a typical early-type galaxy
\citep[e.g.  also][]{mie02,mie04a,mie08c,mis08,mis09}. We double-checked
the radial velocity cross-correlation measurements with an alternative
old stellar population template \citep[NGC\,4636, Schuberth private
communications and][]{sch09}, giving consistent results. Finally,
for those sources with too low cross-correlation amplitude to derive
fiducial radial velocities, the cross-correlation was repeated using
a 'younger' SSP template by \citet{coe07} for an age of 3 Gyrs and
${\rm [Fe/H]}=-0.5~\rm dex$.  This led to the detection of one
additional compact group member in HCG\,22, see Section~\ref{results}.

\begin{figure*}
\begin{center}
\includegraphics[width=17.6cm]{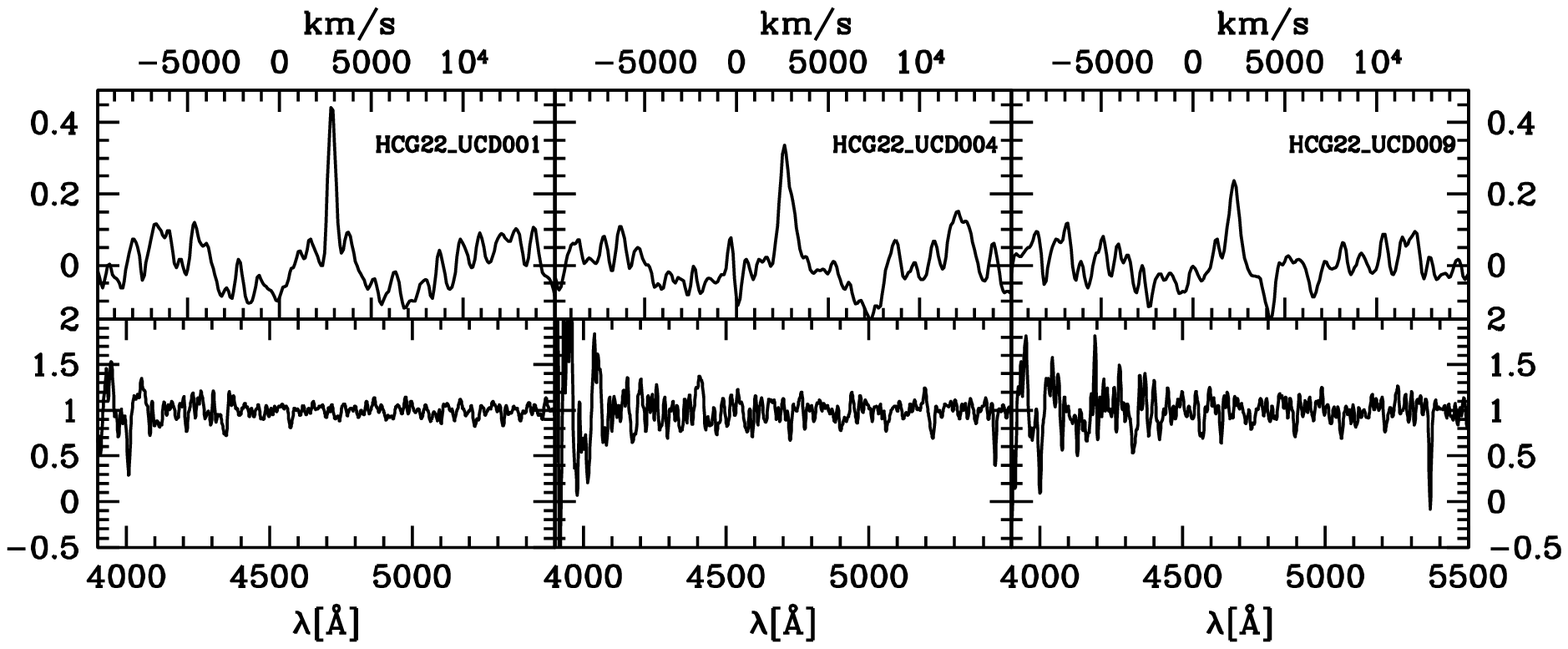}
  \caption{Spectra (bottom panels) and cross-correlation results
  for radial velocity measurement (top panels) for three example
  HCG\,22 compact objects covering our range of cross-correlation
  confidence level (from left to right), see also Table~\ref{table_HCG22}.}
\label{spectra}
\end{center}
\end{figure*}

In Figure~\ref{spectra} we show example spectra and cross-correlation
results for three objects classified as group members according to
their radial velocity.

\subsection{Data reduction: HST Imaging}\label{HST_phot_struct}

We used two data products provided by the HST reduction pipeline -
the $flt$ and the $drz$ images. To avoid problems arising from the
image resampling due to insufficient sub-pixel dithers during the
{\it drizzle} algorithm \citep{fru02} in producing the final $drz$
stack images, we chose to work directly on the $flt$ images for the
actual measurements. The $drz$ images were used for the final
photometric calibrations. The $flt$ images are dark, bias and
flat-field corrected, and have not been resampled, therefore providing
the most natural observation of the astronomical scene.  However,
the ACS $flt$ images are still in the original WFC coordinate system
which suffers from strong geometric distortions.

In order to perform the most accurate structural and point-spread
function-fitting (PSF) photometry measurements of faint, barely
resolved objects one has to properly take into account the varying
pixel area across the ACS/WFC detectors.  \citet{and06} provide an
empirically created library of fiducial PSFs which we used with the
algorithms described in detail in \citet{and00,and06} and \citet{and08}
to perform the PSF-fitting photometry\footnote{The img2xym\_WFC.09x10
code was slightly modified to output the value of the local background
which is important for determining the CTE corrections.}.  PSF-fitting
was performed on every single F475W and F850LP exposure using the
spatially varying array of empirical PSFs, plus a ``perturbation
PSF'' tailored to each exposure to compensate for the temporal
effects of focal variations. At this stage, the catalogs contain
many spurious detections (hot pixels, cosmic rays, and spurious
detections along diffraction spikes). Those were eliminated by
matching the coordinates of each exposure within a tolerance of
0.5\,pixels. The final photometry list contains four measurements
for each source per filter. The final instrumental magnitude is the
goodness of fit (``qfit'') weighted average of the four measurements.

To correct for the charge transfer efficiency (CTE), which is a
function of the date of observation, object brightness, background
and position, we used the latest correction equation provided by
\citet{chi09}.  The photometry was put into the ACS/WFC Vega-mag
system following \citet{bed05}, i.e. by comparing the CTE corrected
instrumental magnitudes with the magnitudes of the same stars
measured within $0\farcs5$ aperture radius and correction from
$0\farcs5$ to ``infinity'' using the values from Table\,1 in
\citet{bed05}. We adopted the zero points from the latest and
improved photometric calibration of the ACS listed on the STScI
website\footnote{Zero points and details on the new calibrations
can be found on http://www.stsci.edu/hst/acs/analysis/zeropoints},
which supersede the previous by \citet{sir05}. We used the ACS/WFC
zero points at $-77\,\rm ^oC$ since the ACS observations were
acquired in May 2006, i.e. before July 2006 when the temperature
of the WFC detector was lowered to $-88\,\rm ^oC$ following the
recovery of the ACS.

The pipeline reduced F450W, F606W and F814W WFPC2 images retrieved
from the HST archive are bias, flat-field and bad pixel corrected
images for each WFPC2 chip. The sub-pixel dithered imaging allowed
us to eliminate cosmic-rays, hot pixels and improve the spatial
resolution.  For image registration we used the IRAF task {\sc
wregister}, with flux conserving drizzle resampling factor of 0.8,
to register the individual exposures for each CCD chip and filter
combination using the solutions recorded in the image headers. This
proved to give an extremely satisfactory final stack image, median
combined from the four exposures with the IRAF task {\sc imcombine}.
We cross checked the profiles of the few stars between the original
exposures and the final median combined image for each filter and
the difference in FWHM was up to $\pm0.02$ pixels. To perform a PSF
photometry on the WFPC2 images, we created a grid of $100\times100$
PSFs with the TinyTim\footnote{TinyTim properly takes into account
the significant variation of the PSF as a function of wavelength
(filter) caused by the expanding of diffraction structures with
increasing wavelength, position-dependent changes of the shape of
the PSF, aberrations, focus offsets between cameras (WFs and PC1
for WFPC2) and wavelength dependent charge diffusion.  See also
http://www.stsci.edu/software/tinytim/} software package \citep{kri95}
for each CCD chip and filter. This library was used to create a
spatially variable PSF model and perform PSF-fitting photometry
with the {\sc allstar} task in IRAF. Isolated stars were used to
derive aperture correction to $1\farcs0$ aperture diameter which
is the aperture used by \citet{dol09} to derive the most up to date
CTE corrections and photometric zero points for each of the WFPC2
chips.

Finally, the ACS and WFPC2 photometry, calibrated to the Vega-magnitude,
was corrected for foreground extinction estimated from the \citet{sch98}
dust maps toward the direction of both groups. The estimated Galactic
extinction is $E(B-V)=0.026$ and $0.055~\rm mag$ for HCG\,90 and
22, respectively. We used the \citet{car89} relations and $R_V=3.1$
to calculate the absorption at the effective wavelengths for the
ACS/WFC and WFPC2 filters. Thus the following dereddening values
were applied for the HCG\,90 ACS images: $A_{F475W}=0.104$,
$A_{F850LP}=0.039~\rm mag$ and for the HCG\,22 WFPC2 images:
$A_{F450W}=0.237$, $A_{F606W}=0.170$ and $A_{F814W}=0.112~\rm mag$.
The result is shown in the colour-magnitude and colour-colour
diagrams (Figures~\ref{Fig:H90_CMD} and~\ref{Fig:H22_CMD_CCD}) and
discussed in Section~\ref{Sect:HST-Sizes-colors}.

\subsubsection{Measuring structural parameters of UCD candidates}

Measuring structural parameters of UCD candidates in HCG\,22 and
HCG\,90 is feasible thanks to the supreme resolution of the ACS and
WFPC2 cameras of $0\farcs05$ and $0\farcs1$ pixel$^{-1}$, corresponding
to $\sim8$ and $16$ parsecs at a distance of $33.1~\rm Mpc$,
respectively. Therefore, and due to the very well characterized PSF
over the ACS/WFPC2's field-of-view, extended objects with half-light
radius of only a few parsec will be measurable, allowing to resolve
objects down to typical GC sizes. To derive the $r_h$ of the UCD
candidates in our sample from the ACS images, we generated a ten
times subsampled PSF from the \citet{and06} ACS/WFC ePSF library.
Each PSF is tailored to the position on the chip for every object
measured. This PSF was used with {\sc ishape} of the {\sc baolab}
software package\footnote{We used the most recent release of
\protect{\sc baolab} found on http://www.astro.uu.nl/~larsen/baolab/}
\citep{lar99} which models the object profile with an analytical
function convolved with a (model) PSF.  We modeled all objects with
King profiles with concentrations of tidal-to-core radius of
$r_t/r_c=5, 15, 30$ and 100 and adopted the structural parameter
measurements from the best $\chi^2$ fit model.  The output is the
FWHM along the object semi-major axis. It is converted to $R_{\rm
eff}$ using the coefficients relating those two values, as prescribed
in the {\sc baolab} manual.  This $R_{\rm eff}$ needs to be corrected
for object ellipticity and brought to the geometrically mean value
(``effective'' $r_h$) by multiplying by the square root of the
major/minor axis ratio \citep[details see eq.\,1 in][]{geo08}.

Measurements of the structural parameters of UCD candidates on the
WFPC2 images were performed in an analogous way as for the ACS, but
PSFs at the objects' position were generated with the TinyTim for
the F606W filter. Generating ten times subsampled PSFs with TinyTim
does not include a convolution with charge diffusion kernel (CDK),
which additionally smears the stellar PSF. The CDK available within
TinyTim is applicable in the 500 - 600 nm range (F555W filter),
therefore, we restricted our measurements only to the F606W filter
and did not perform measurements for the rest of the filters.

We finally note that reliable $r_{\rm h}$ measurements with {\sc
ishape} are performed for objects with $S/N>20$ \citep{lar99}. The
UCD candidates in the HCG\,22 have signal-to-noise in the range
$20<S/N<80$ and the UCD candidate in HCG\,90 has $S/N=86$.

\section{Results}
\label{results}

\begin{figure*}
\begin{center}
\includegraphics[width=8.6cm]{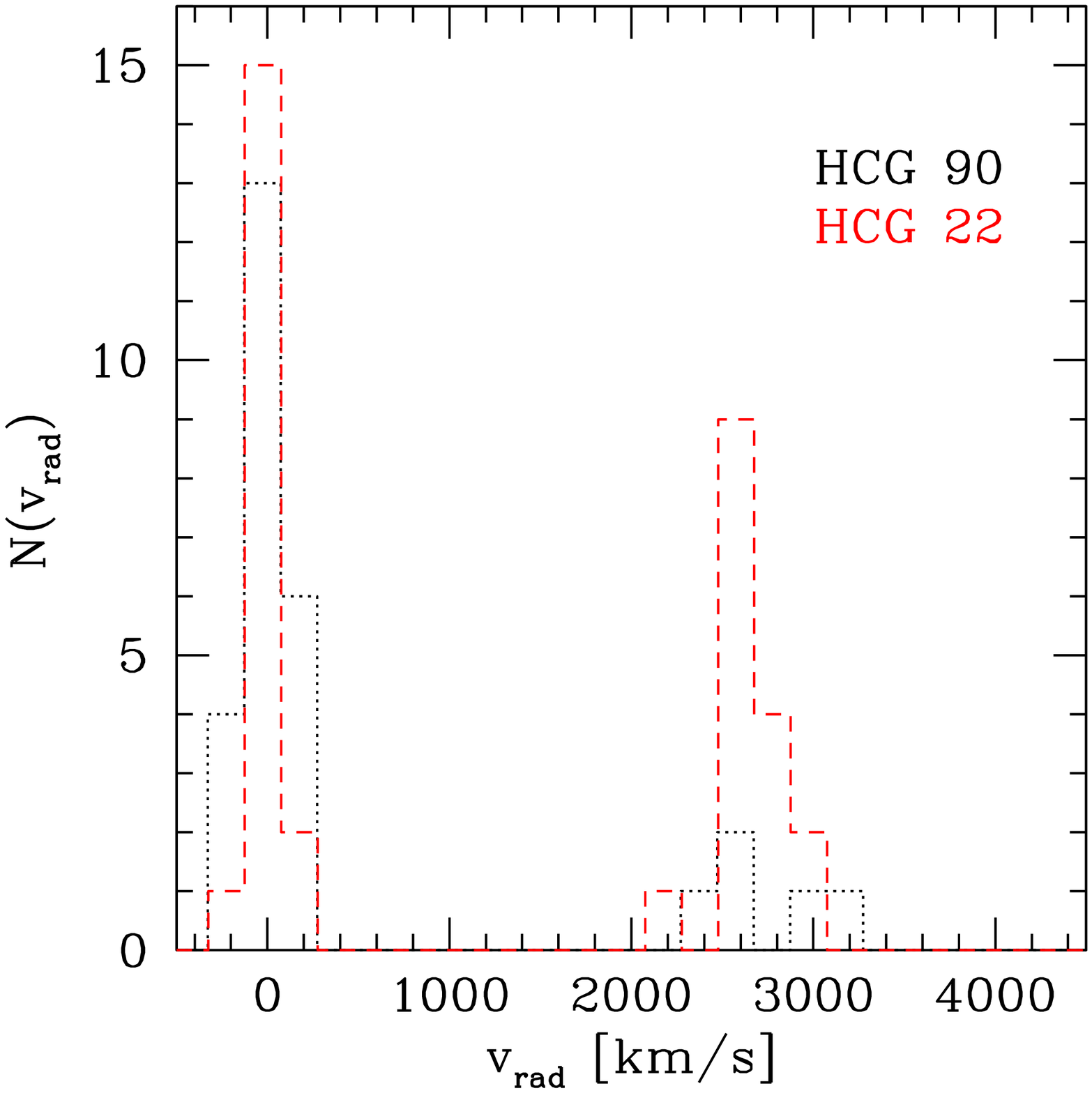}
\includegraphics[width=8.6cm]{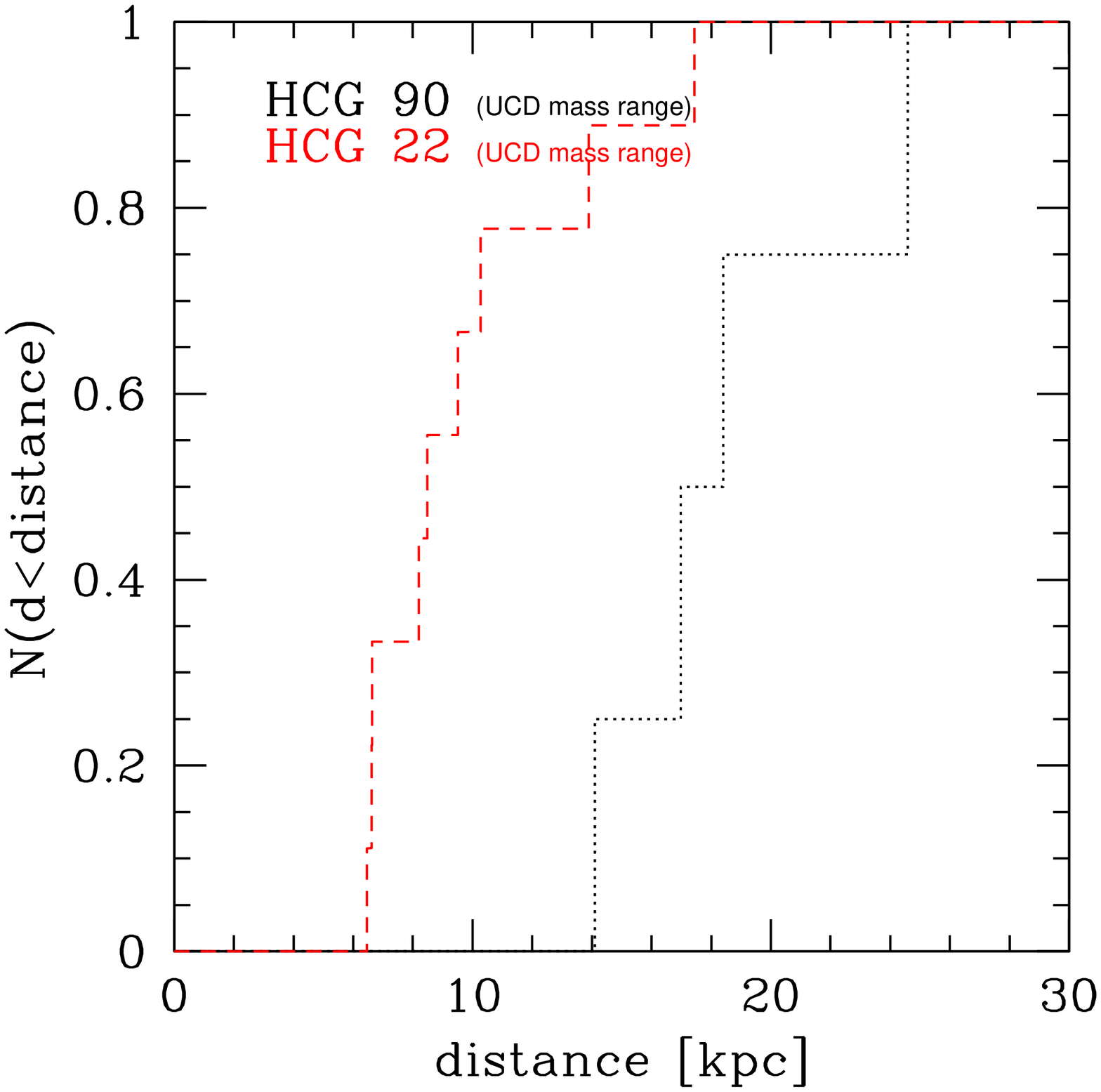}
  \caption{{\bf Left panel:} Radial velocity histogram (corrected
  to heliocentric velocity) of spectroscopic targets in HCG\,22
  (red)  and HCG\,90 (black). The group members at around $2700~\rm
  km~s^{-1}$ are well separated from foreground stars and background
  galaxies with extreme velocities (not shown for clarity). {\bf
  Right panel:} Cumulative radial distance distribution of the HCG
  compact members relative to the central galaxies (NGC\,1199 and
  NGC\,7173 for HCG\,22 and 90, respectively). The samples in this
  panel are restricted to the UCD mass range. The KS-test probability
  of both distributions having the same parent distribution is
  0.9\%.
  }
\label{vradhist}
\end{center}
\end{figure*}

In Figure~\ref{vradhist} we show a histogram of the measured radial
velocities (corrected to heliocentric velocity). We detect five
members of HCG\,90 and 16 members of HCG\,22. Four other sources
were found to be background galaxies at significantly higher redshift
than the HCGs. All other targets turned out to be foreground stars.
We note that one of the 16 members in HCG\,22 (HCG22\_UCD011) was
detected only via cross-correlation with the intermediate age (3
Gyr) SSP template \citep{coe07}, instead of the reference old stellar
population template.  Besides for excluding very young objects, our
selection criteria were satisfactorily efficient, in the UCD richer
group (HCG\,22), more than 50\% of the high priority targets we
indeed objects belonging to the group.

The properties of all compact objects are summarized in
Tables~\ref{table_HCG22} to~\ref{table_HCG90_HST}.  They span
absolute magnitude ranges of $-10.4>M_R>-11.1~\rm mag$ for HCG\,90
and $-10.0>M_R>-11.5~\rm mag$ for HCG\,22.  Their positions within
the groups are shown with large open (red) squares on the pre-images
in Figure~\ref{images}. The HCG\,22 UCD candidates are clearly
clustered around the group dominant galaxy NGC\,1199, while the
distribution in HCG\,90 is more uniform over the group area.

Their $R$ vs. $(B-R)_0$ colour-magnitude distribution are shown in
Figure~\ref{CMD}. In the left panel of Figure~\ref{CMD} are shown
model predictions from \citet{bru03} in the colour-magnitude range
of objects at the approximate mass limit of $2\times10^6 M_{\odot}$
between normal GCs and UCDs, for a range of metallicities and ages.
Assuming a 12 Gyr stellar population, 9 of the 16 objects in HCG\,22
are in the UCD mass range range, with $M_R \lesssim -10.4~\rm mag$.
For HCG\,90, four of the five objects are in the UCD mass range
range.  The colour range $1.1<(B-R)_0<1.65~\rm mag$ of the HCG
compact members is consistent with them being old and covering a
range of metallicities $-1.7 \lesssim {\rm [Fe/H]} \lesssim 0.2~\rm
dex$. We note, however, that moderately metal-rich and young objects
of a few Gyr would also still fall in the observed colour window.

In the colour-magnitude diagrams (CMDs) in Figure~\ref{CMD} we also
indicate the data of all candidate compact objects, and all objects
that were observed with the MXU masks. The group member detections,
indicated as filled circles, concentrate towards the faint end of
our survey. We have not found very massive UCD candidates with
$M_R\simeq -14.0$ ($\simeq 10^8 M_{\odot}$) as seen in the Fornax
and Virgo galaxy clusters, nor presumably young UCDs among the blue
candidates in the covered colour range.

\vspace{0.5cm}
\noindent In Section~\ref{Sect:HST-Sizes-colors}, we present the
sizes and colours derived for a sub-sample of UCD candidates from
HST imaging. In Section~\ref{Sect:Kinematics-spatial-distribution}
we analyze the kinematics and spatial distribution of the UCD
candidates, and in Section~\ref{Sect:stellar-populations} we discuss
their stellar populations.

\subsection{Sizes and colours from the HST imaging data}
\label{Sect:HST-Sizes-colors}

\begin{figure}
\includegraphics[width=0.5\textwidth, bb=70 70 410 300]{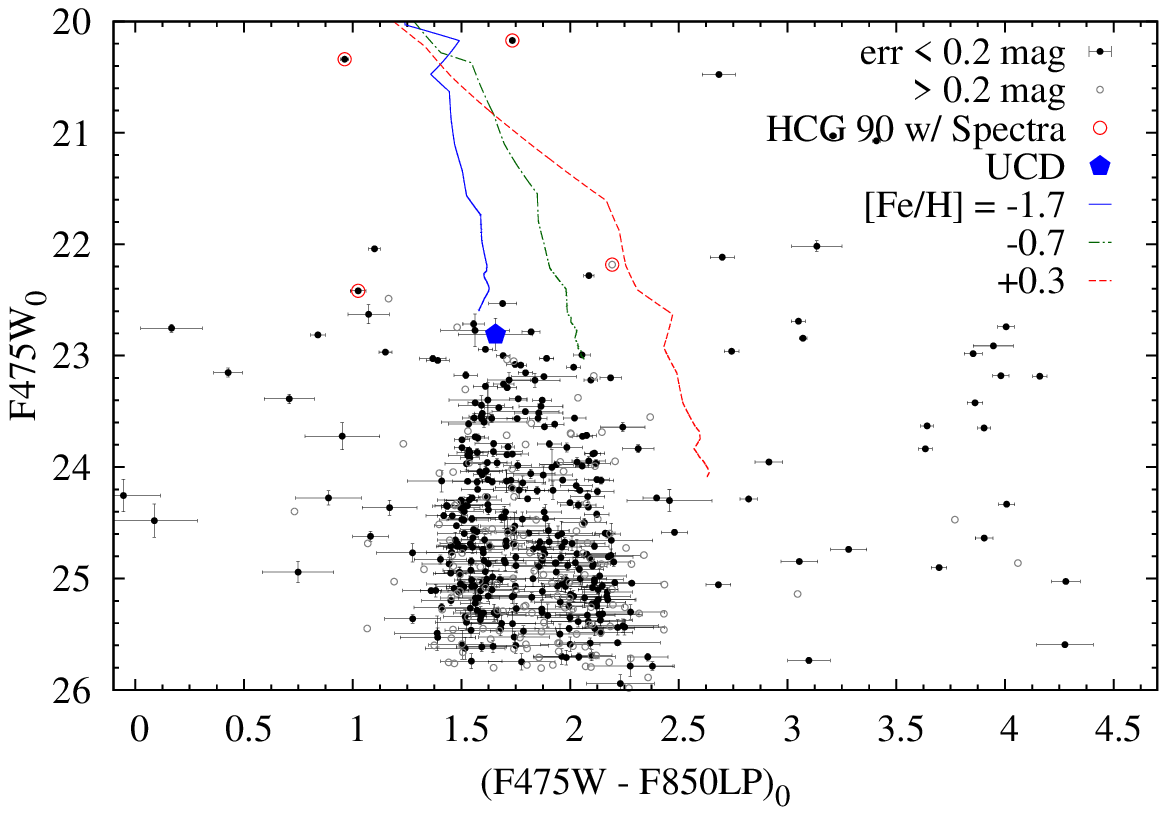}
\caption{Colour-magnitude diagram of all sources in the HCG\,90
WFC/ACS field of view. The UCD candidate is indicated with a solid
pentagon while large open circles indicate all sources for which
spectroscopy is available. For comparison, GALEV SSP model
isometallicity tracks are shown with lines for three fiducial
metallicities \citep{and03,kot09}, as indicated in the figure's
legend. The models luminosities are for a cluster of
$5\times10^6M_{\odot}$.
\label{Fig:H90_CMD}
}
\end{figure}

\begin{figure*}
\includegraphics[width=1.0\textwidth]{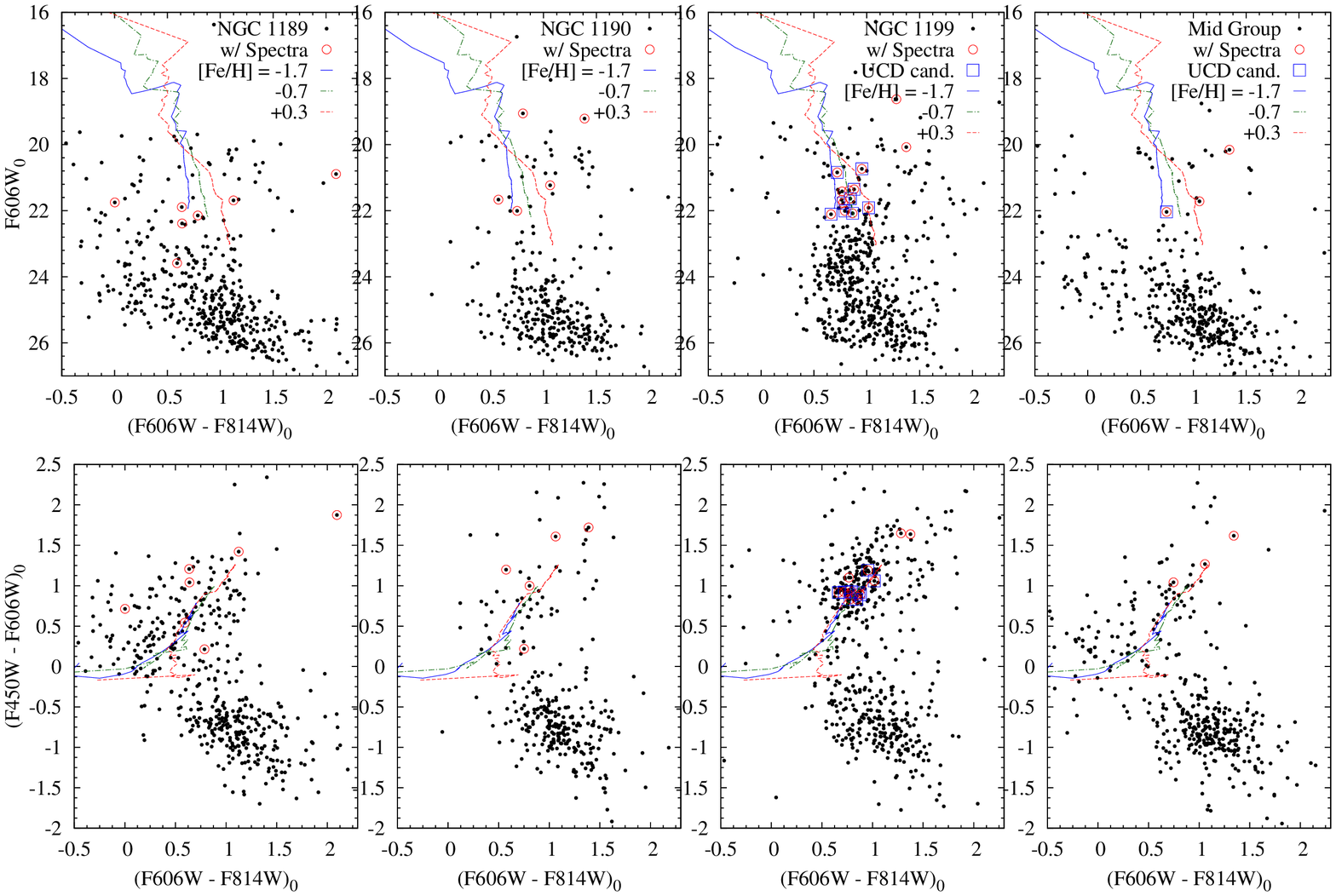}
\caption{{\bf Top row:} Colour-magnitude diagrams for all objects
(dots) in HCG\,22 for each of the four WFPC2 pointings. The
legend for each CMD indicates the galaxy/region at which the WF3
chip was centered. {\bf Bottom row:} Colour-colour diagrams of all
sources (dots) with photometry for the same fields. In all panels
the open circles indicate the objects with spectroscopy while the
open squares are the UCD candidates. The description of the SSP
isometallicity tracks is the same as in Figure~\ref{Fig:H90_CMD}.
\label{Fig:H22_CMD_CCD}
}
\end{figure*}

\begin{figure}
\includegraphics[width=0.5\textwidth]{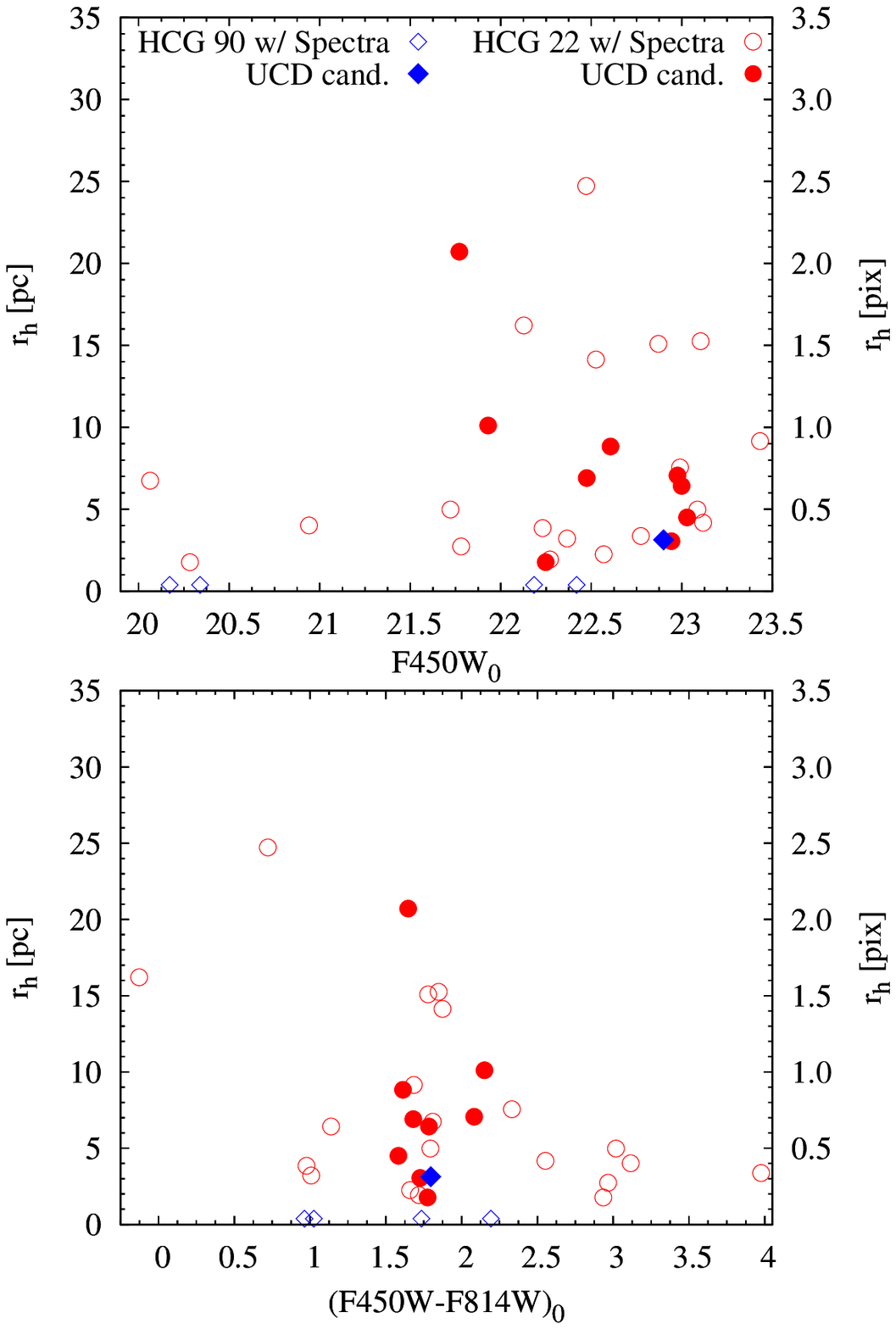}
\caption{Distribution of the half-light radius in parsecs (left
axis) and pixels of $0\farcs1$ (right axis) versus the luminosity
(top) and colour (bottom) of objects with spectroscopy in HCG\,90
and 22. Blue pentagons show objects with spectra in HCG\,90 and red
circles in HCG\,22. Solid symbols mark the UCD candidates in both
groups. Unresolved objects in this figure will have sizes close to
$r_{\rm h}=0$. Pixel to parsec conversion in both panels corresponds
to the WFPC2 pixel scale ($0\farcs1~\rm pix^{-1}$). Thus, HCG\,90
objects will have twice the pixel value on the y-axis, because of
the higher resolution of the ACS/WFC ($0\farcs05~\rm pix^{-1}$).
For comparison purpose, ACS (F475W-F850LP) colours were transformed
to the WFPC2 (F450W-F814W) using GALEV \citep{and03,kot09} for the
metallicity range of our objects.
\label{Fig:H22_H90_Structpar}
}
\end{figure}

Important constraints on the nature of the detected UCD candidates
can be obtained from the high resolution HST imaging. In
Figures~\ref{Fig:H90_CMD} and \ref{Fig:H22_CMD_CCD} we show the
colour and magnitude distributions of all measured objects in the
ACS/WFC and WFPC2 fields of view. In Figure~\ref{Fig:H22_H90_Structpar}
are shown the half-light radius measurement as a function of object
luminosity and colour for all objects with spectroscopy in our
sample. Here we focus our discussion on the properties of the UCD
candidates preselected from the spectroscopic sample.

All UCD candidates are consistent with the expectations from simple
stellar population (SSP) evolutionary models
\citep[GALEV\footnote{http://www.galev.org/} by][]{and03,kot09} for
old ages and the entire range in metallicities. The comparison in
Figure~\ref{Fig:H90_CMD} between the only UCD candidate (HCG90\_UCD005)
within the ACS field of HCG\,90 and the SSP model with a mass
of $5 \times 10^6 M_{\odot}$, shows that the UCD candidate can be
as massive as $\sim10^7 M_{\odot}$. This is consistent with the
estimate obtained from the ground-based photometry in Figure~\ref{CMD},
indicating that HCG90\_UCD005 falls in the transition mass regime
between UCDs and GCs. Its light profile is best fit by a King model
with concentration $c=r_t/r_c=100$ and a half-light radius of $r_{\rm
h}=3.13~\rm pc$.  The $r_h$ is typical for GCs, and thus HCG90\_UCD005
morphologically classifies rather as a star cluster than as a UCD.
Unfortunately, it has too low S/N spectrum for quantitative abundance
analysis. This object is indicated by a solid pentagon in
Figure~\ref{Fig:H22_H90_Structpar}.  The other brighter sources
with spectra (circles in Figure~\ref{Fig:H90_CMD}) are foreground
stars which is further confirmed by their $r_{\rm h}\simeq0~\rm
pc$, i.e. their PSF is indistinguishable from a stellar PSF (cf.
Figure~\ref{Fig:H22_H90_Structpar}).

Figure~\ref{Fig:H22_CMD_CCD} shows the colour-magnitude and
colour-colour diagrams of sources in the four HST/WFPC2 pointings
covering most of HCG\,22.  We have labeled these fields with the
name of the target at which the WF3 chip was centered. These are
the galaxies in this group (the brightest NGC\,1199, NGC\,1189 and
NGC\,1190) and one pointing in between them. Ten of the 16 HCG\,22
UCD candidates are within the WFPC2 fields. Nine of those are within
the NGC\,1199 pointing, the other in the mid-group pointing.  All
of them are brighter than the $5\times10^6M_{\odot}$ SSP models
(see also Figure~\ref{CMD}), thus falling into the mass range of
UCDs.  The most luminous UCD candidate has an inferred mass of
$\simeq 10^7 M_{\odot}$. We note that the dynamical M/L ratios
of UCDs in other environments have been found to be 50-100\% above
predictions from SSP models \citep[e.g.][]{mie08c,dab08}, and are
on average around $M/L=4-5$.  Adopting this average M/L ratio instead
of $M/L \sim 3$ characteristic for the SSP models in
Figure~\ref{Fig:H90_CMD} and also typical for the most metal-rich
Galactic GCs, the implied mass range of our UCD candidates would
extend to about 50\% larger values up to $\sim 2\times10^7 M_{\odot}$.

The candidates have a broad range of half-light radii $2<r_{\rm
h}<21~\rm pc$ with a mean of $r_{\rm h}=7.4\pm1.5~\rm pc$. That is,
they are on average 2-3 times larger than typical GCs and up to
four times more massive than $\omega$Cen, hence they can be classified
as UCDs. It is interesting to note the apparent concentration of
blue objects on the NGC\,1189 and ``Mid Group'' colour-colour
diagrams (bottom row the left and right most panels).  NGC\,1189,
covered by the ``Mid Group'' WFPC2 field, is a face-on, active
star-forming SBcd which is the group brightest galaxy in UV wavelengths
\citep{tza10}. Therefore, young objects, like unresolved star-forming
regions, young star clusters are expected to be present in our
color-magnitude diagrams.  NGC\,1190 and 1199 are considered as
passive galaxies and the absence of blue (young) sources is expected.

The photometric and structural properties of the observed UCD
candidates are summarized in Tables~\ref{table_HCG22_HST}
and~\ref{table_HCG90_HST}.

\subsection{Kinematics and spatial distribution}
\label{Sect:Kinematics-spatial-distribution}

For HCG\,22, the mean velocity and dispersion of the 16 HCG compact
members are $v_{\rm mean,HCG\,22}=2664 \pm 47~\rm km~s^{-1}$ and
$\sigma_{\rm HCG\,22}=171_{-40}^{+71}~\rm km~s^{-1}$, where $\sigma$
was corrected for internal errors using \citet{dan80} and the stated
errors correspond to a 90\% confidence interval.  For HCG\,90, the
values are $v_{\rm mean,HCG\,90}=2608 \pm 109~\rm km~s^{-1}$ and
$\sigma_{\rm HCG\,90}=206_{-80}^{+405}~\rm km~s^{-1}$.  The mean
radial velocities are in good agreement with the group values,
$v_{\rm mean,HCG\,22}=2629\pm33~\rm km~s^{-1}$ and $v_{\rm
mean,HCG\,90}=2545\pm58~\rm km~s^{-1}$ \citep{rib98,zab98a}.  The
velocity dispersions for the UCD candidates in HCG\,90 agrees well
with the group value \citep[$193_{-33}^{+36}~\rm km~s^{-1}$][]{zab98a}
showing no kinematical decoupling from the overall galaxy population,
while the dispersions for candidates in HCG\,22 is considerably
higher than the group value \citep[$40\pm28~\rm km~s^{-1}$][]{rib98},
being more consistent with the internal velocity dispersion of
NGC\,1199 \citep[$207\pm21~\rm km~s^{-1}$][]{pru96}.

In HCG\,22, we detect about twice as many UCD candidates as in
HCG\,90 (9 vs. 4), but taking into account the lower completeness
for HCG\,90 (76\% for HCG\,22 and 49\& for HCG\,90), this difference
is not significant. Figure~\ref{images} shows that the compact
objects in HCG\,22 appear to be more strongly clustered around the
group's central galaxy than is the case for HCG\,90, a possible
evidence of their origin being associated to processes related to
its central galaxy (NGC\,1199). In Figure~\ref{vradhist} right
panel, we compare the cumulative radial distance distribution of
UCD candidates in HCG\,22 and HCG\,90 with respect to the central
galaxy (NGC\,1199 and NGC\,7173 for HCG\,22 and 90, respectively).
Applying a Kolmogorov-Smirnov-test (KS test), the probability that
the HCG\,22 and HCG\,90 radial distance distributions are drawn
from the same parent population is only 0.9\%. This result confirms
the visual impression of a stronger clustering in HCG\,22.  We note
that our magnitude cut for excluding objects in the GC magnitude
range excludes the HCG\,90 source with closest projected distance
to the central galaxy.  Including this object into the KS test still
yields a probability of only 4\%.

\subsection{Stellar populations}

\label{Sect:stellar-populations}

We have measured Lick line indices for all the compact group members
detected, using the passband definitions from \citet{tra98}, having
smoothed the spectra to match the 9 {\AA} Lick system resolution.
From measurements of six Lick standard stars taken in the same
setting and smoothed in the same way, we could not detect any
significant deviation between tabulated Lick line index values and
our measurements down to a level of $\sim$0.1 {\AA}. Given the much
larger measurement errors of our science targets (see below) we
therefore do not apply any correction.

A commonly used metallicity sensitive index is $<Fe>=0.5 \times
({\rm Fe5270} + {\rm Fe5335})$. In the following, we only show and
discuss measurements for which the statistical uncertainty of the
$<Fe>$ index is less than 0.5 dex, as determined from the photon
counts in the object and sky spectra. Tables~\ref{table_HCG22} and
~\ref{table_HCG90} show the line index estimates for the five sources
in HCG\,22 and three sources in HCG\,90 with errors below 0.5 dex.

We also indicate an estimate of the corresponding ${\rm [Fe/H]}$.
For this we use for ${\rm [Fe/H]}<-0.4~\rm dex$ the empirical
calibration between ${\rm [Fe/H]}$ and ${\rm [MgFe]}$ derived by
\citet{puz02} from galactic GCs. For ${\rm [Fe/H]}>-0.4~\rm dex$
we adopt an extrapolation of the form ${\rm [Fe/H]}_*=a+b \ \times
\ \log~(\rm MgFe)$ that matches the calibration for ${\rm
[Fe/H]}<-0.4~\rm dex$ \citep[see also][]{mie07b}.

\begin{figure*}
\begin{center}
\includegraphics[width=8.6cm]{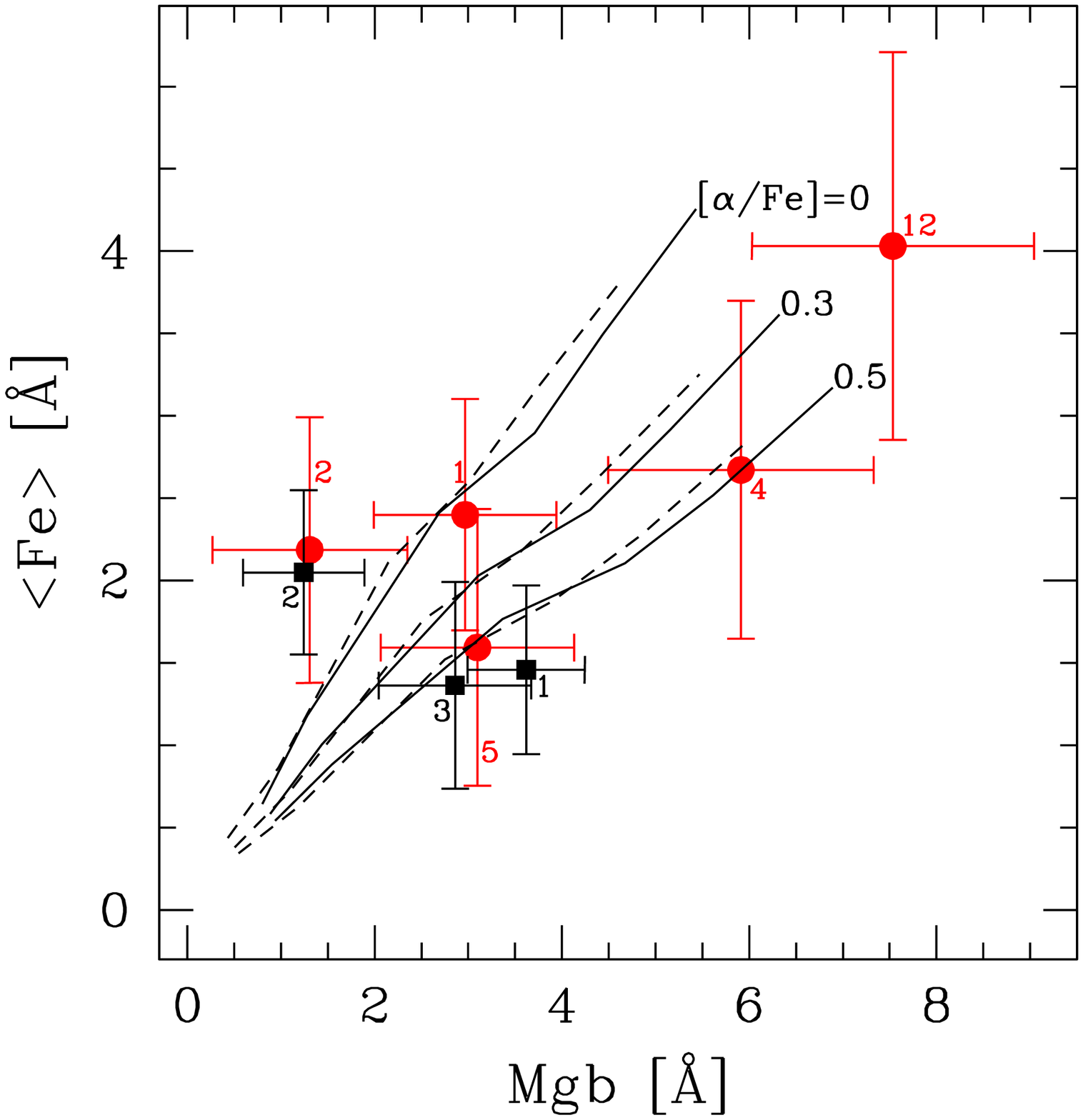}
\includegraphics[width=8.6cm]{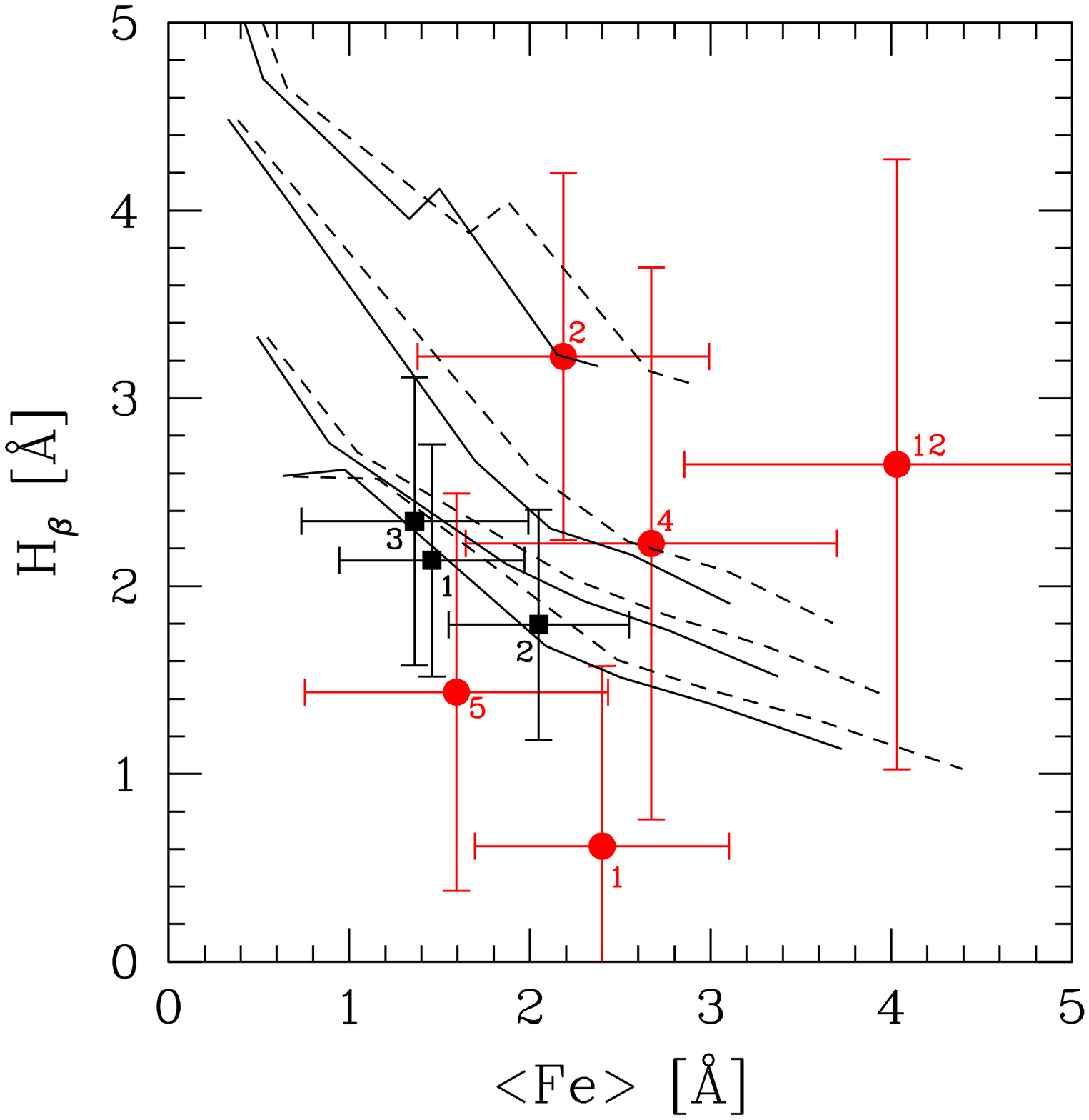}
  \caption{Lick line index diagnostic plots for a sub-sample of the
  compact group members with high enough S/N (error in $<Fe>$ index
  below 0.5 dex). {\bf Left panel:} Diagnostic plot for alpha
  abundances, plotting $<$Fe$>$ vs. Mgb index. We show lines of
  identical alpha abundances for both age of 12 Gyr (solid line)
  and 4 Gyr (dashed line), using the model predictions of \citet{tho03}.
  The two most metal-rich members of either group appear to have
  super-solar alpha abundances. {\bf Right panel:} Diagnostic plot
  for age estimation, plotting $H_{\beta}$ vs $<$Fe$>$ index for
  four different ages 12, 7, 3, and 1 Gyr from bottom to top (solar
  $\alpha-$abundance). The three UCDs in HCG\,90 appear consistent
  with very old ages, as do the two HCG\,22 UCDs HCG22\_UCD001 and
  HCG22\_UCD005 (see Tables~\ref{table_HCG22} and~\ref{table_HCG90}).
  Numbers on the symbols are the ID of each object, where the
  prefixes were omitted for clarity.}
\label{Lick}
\end{center}
\end{figure*}

In Figure~\ref{Lick} we show diagnostic plots for age and alpha
abundances, using model predictions from \citet{tho03}.  From the
left panel it can be seen that the two most metal-rich UCD candidates
in either groups, appear most consistent with super-solar alpha
abundances of 0.3 to 0.5 dex. Furthermore, in the age diagnostic
plot in the right panel, it is clear that all three UCD candidates
in HCG\,90 are most consistent with old ages of around a Hubble
time. For HCG\,22, object HCG22\_UCD002 is inconsistent with an old
age of 12 Gyr, indicating the presence of an intermediate age stellar
population.  The other sources have too large error bars to make
strong statements.

\begin{figure}
\begin{center}
\includegraphics[width=8.6cm]{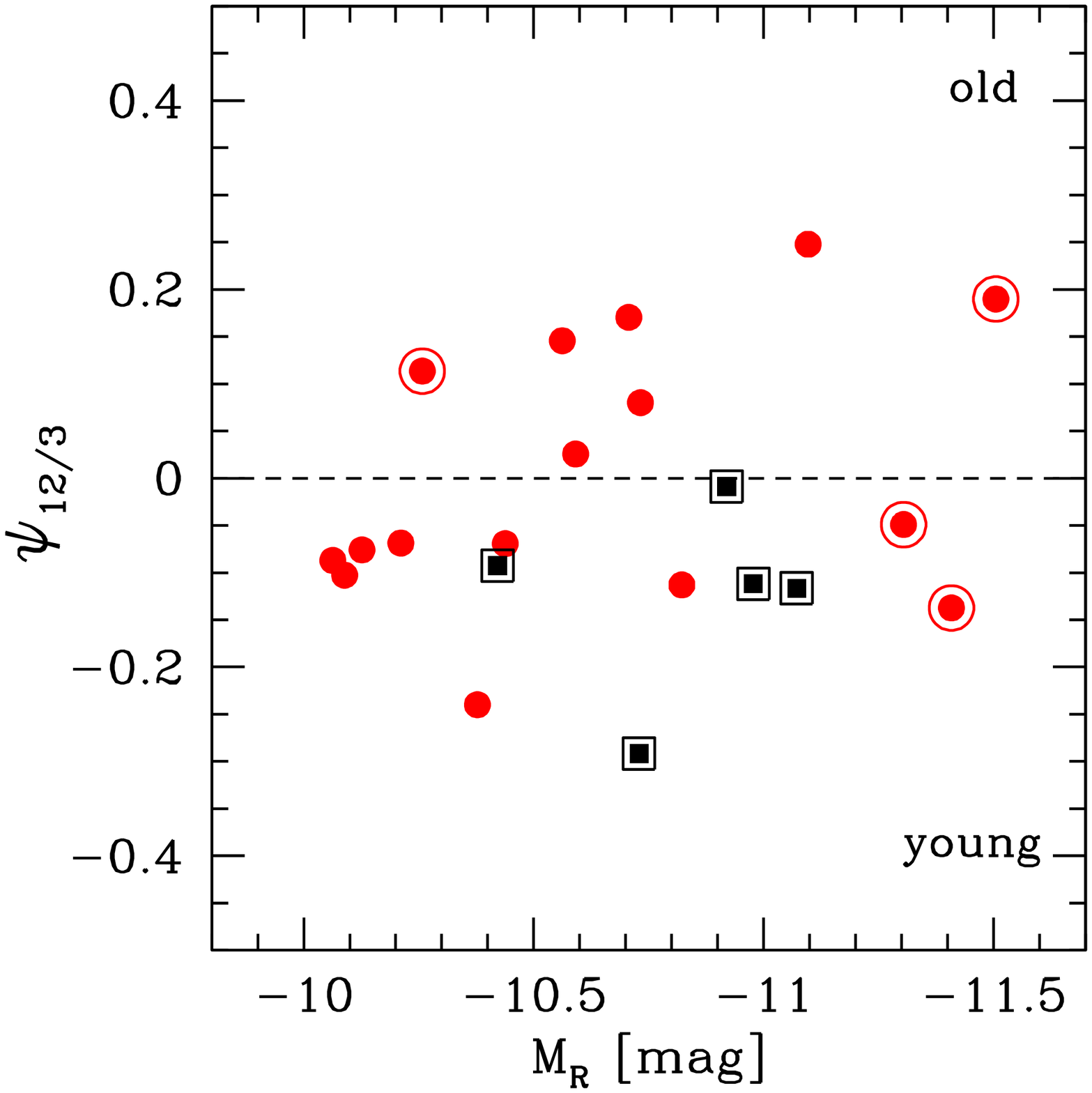}
\caption{Absolute magnitude of confirmed cluster members (HCG\,22
are filled red circles, HCG\,90 are black squares) vs. the ratio
between the cross-correlation amplitude using our reference old
stellar template and an intermediate age template \citep[3 Gyrs,
${\rm [Fe/H]}=-0.5~\rm dex$;][]{coe07}. Negative to positive values
indicate spectra better fit by young to old ages, respectively.
Objects marked are the ones with cross-correlation confidence level
higher than 5.}
\label{2templates}
\end{center}
\end{figure}

To further investigate the possibility of intermediate age UCDs,
we show in Figure~\ref{2templates} the ratio of the cross-correlation
amplitude of our reference old stellar template \citep[see
also][]{mie02,mie04a,mie08c,mis08,mis09}, and the intermediate age
stellar template of 3 Gyrs and ${\rm [Fe/H]}=-0.5~\rm dex$ dex from
\citet{coe07}. We define this ratio as $\psi_{12/3}=\log(\rm
XcorAmp(12Gyr)/XcorAmp(3Gyr))$. Sources which are better fit by the
old stellar template lie above the (dashed) identity line, while
sources fit better by the intermediate age template lie below the
line. There are two UCD candidates, HCG22\_UCD011 and HCG90\_UCD004,
which are distinctively better fit by the intermediate age stellar
template. Those two sources are good candidates for intermediate
age UCDs.  Unfortunately, the spectra were not of sufficiently high
S/N to investigate this possibility via line index measurements.
It is noteworthy that HCG22\_UCD002 is, after HCG22\_UCD011, the
source with the second best intermediate age fit in HCG\,22.  This
supports the indications from line index measurements that this
object also contains intermediate age stellar populations.  On
average, UCDs in HCG\,90 are fit slightly better by the intermediate
age stellar template than UCDs in HCG22 (cf. Figure~\ref{2templates}).

\section{Discussion and Conclusion}
\label{conclusions}

\subsection{Summary of the results}

Before proceeding with the discussion, in the following we present
a summary of the main results.

(1) We detected 16 and 5 objects belonging to HCG\,22 and HCG\,90,
respectively. They cover the magnitude range $-10.0>M_R>-11.5$ mag
for HCG\,22 and $-10.4>M_R>-11.1$ mag for HCG\,90. Their colours
are consistent with old stellar populations covering a broad range
of metallicities. Metallicity estimates from line index measurements
for a sub-sample of objects with sufficiently high S/N cover a range
of $-1.0 \lesssim {\rm [Fe/H]} \lesssim 0.3~\rm dex$.

(2) Photometric mass estimates using \citet{bru03} put four objects
in HCG\,90 and nine objects in HCG\,22 in the mass range of UCDs ($>
2\times10^6 M_{\odot}$) for an assumed age of 12 Gyr.

(3) The detected objects are on average 2-3 times larger than the
average size of Galactic GCs, covering a range of $2\lesssim
r_{h}\lesssim 21~\rm pc$.

(4) The mean velocities of the UCDs agree with the galaxy group
values.  The velocity dispersion of HCG\,90 UCDs is similar to that
of the galaxies in the group, while the velocity dispersion of
HCG\,22 UCDs is more consistent with the internal dispersion of
NGC\,1199.

(5) The UCDs in HCG\,22 are more concentrated around the central
galaxy than in HCG\,90, at the 99\% confidence level.

(6) We find two UCDs in HCG\,22 and one in HCG\,90 with evidence
for an intermediate age stellar population (3-5 Gyrs), based on
spectral cross-correlation with old and intermediate age templates
and Lick line index measurements.  The clearest example for this
is the largest and most massive UCD ($\sim 10^7 M_{\odot}$) in our
sample, detected in HCG\,22.

\subsection{The newly detected UCDs in the group context}

The two investigated Hickson Compact Groups present different
evolutionary stages. While HCG\,22 has no evidence of intragroup
light, HCG\,90 has about a third of its total light in the IGL
component (Da Rocha \& Ziegler 2010 - in preparation). Also the
X-ray properties of the groups are different. \citet{pon96} established
an upper limit for the ROSAT detection of HCG\,22 and \citet{whi00}
detected NGC\,1199, presumably the group center, around which the
UCDs are concentrated, as a point source on a new analysis of ROSAT
data.  In HCG\,90 a diffuse X-ray component was detected, which
shows strong emission on the interacting pair NGC\,7174/7176 and
is in good agreement with the space distribution of the IGL component,
in addition to individual emission from the member galaxies
\citep{pon96,mul98,whi03,osm04,tre05}.

The spatial distribution of UCDs in each group is not discrepant
with the respective distribution of the overall light (and stellar
mass) in each group. HCG\,22 is a very concentrated structure with
three galaxies and NGC\,1199 represents about 75\% of the group's
luminosity. Hence the concentration of UCDs around NGC\,1199 is not
too surprising, since even a component associated to the group's
potential would mainly be distributed around this galaxy, as is its
GCS \citep{dar02}.  In contrast, HCG\,90 has a much more uniform
luminosity distribution, where the difference between the brightest
and the faintest galaxy is less than 10\% of the total group
luminosity and the brightest component is the IGL, in which the
three studied galaxies and the UCDs are embedded.  The more uniform
mass/light distribution is reflected by the spatial distribution
of the UCDs found in this group.

How does the specific frequency of UCDs (number of UCDs normalized
by the total light of the group in the $B$-band) in either group
compare? This quantity can be useful to compare the richness and/or
formation efficiency of UCDs in different galaxy groups and clusters.
We define the UCD specific frequency normalized to $M_B=-20~{\rm
mag}$, hence $S_{N,{\rm UCD}}=N_{\rm UCD} \times 10^{0.4 \times
(M_B+20)}$. This normalization results in UCD specific frequencies
of comparable numerical values to GC specific frequencies, yielding
for example $S_{N,{\rm UCD}} \simeq 1$ for the case of a single UCD
in an L$_*$ galaxy. For our purposes we need to restrict the group
luminosity estimate in HCG\,90 to the 3 of the 4 group galaxies
covered by our survey, but include the diffuse intragroup light
fraction of $\sim$35\% which was not taken into account by
\citet{hic89}.  Assuming the same distance modulus $(m-M)=32.6~{\rm
mag}$ for both groups, we find a total $B$-band luminosity of
$M_B=-21.5$ mag for HCG\,90 within our survey area, and $M_B=-20.7$
mag for HCG\,22 \citep{hic89}. Taking into account the number of 4
and 9 UCDs found in either group and the corresponding survey
incompleteness, we get $S_{N,{\rm UCD,HCG\,90}}=2.0 \pm 1.0$ and
$S_{N,{\rm UCD,HCG\,22}}=6.3 \pm 2.1$.  This difference of about a
factor of three indicates that the formation process of UCDs has
been more efficient in HCG\,22 than in HCG\,90.

We can extrapolate the GC luminosity function (GCLF) estimated for
NGC\,1199 in HCG\,22 \citep{dar02} to the magnitude range investigated
by us to get an idea how many GCs we should expect to detect. The
GCLF of NGC\,1199 can be described by a Gaussian with a turnover
magnitude of $M_V = -7.33$ and a dispersion of $\sigma = 1.40$
\citep{har01}, and the whole GCS has an estimated total number of
objects of $314\pm105$.  With our survey we probe about 6.7\% of
the GCLF and that corresponds to $16\pm4$ objects in the observed
area. This matches the number of 16 compact stellar systems we have
detected (21 if corrected for incompleteness) in HCG\,22.  Hence,
we do not detect an overabundance of bright compact objects with
respect to an extrapolated GCLF. The magnitude of the brightest UCD
detected in either group ($M_V\simeq -11$ mag) is typical for the
luminosity of the brightest compact object found in galaxies that
fall in the same luminosity range $-21 \lesssim M_V \lesssim -20$
mag as the ones investigated here \citep{hil08}. In the Local Group,
GCs/UCDs brighter than $M_V = -9.5$ (approximately the faint limit
for our sample) are of comparable abundance to that found in the
two HCGs: there are two objects in the Galaxy \citep{har96} and
between five and ten objects in M31 \citep{bar00} depending on the
distance modulus adopted. Even though the studied groups are about
twice as large as the Local Group, given the uncertainties, the
number of bright GCs/UCDs is of the same order.

The abundance of ``alpha elements'' ([$\alpha$/Fe]) is an indicator
of the star formation time scales and can help to distinguish between
different formation channels of UCDs.  GCs, on average, have
super-solar [$\alpha$/Fe], being ``alpha enhanced''
\citep{puz06,col09,woo10,tay10}, reflecting their short formation
time-scales. UCDs present a broader range of [$\alpha$/Fe] abundances
which varies as a function of environment \citep{mie07b,hau09}. In
Figure~\ref{Lick} it is shown that from our eight UCDs with Lick
index measurements, one in each HCG presents slightly sub-solar
[$\alpha$/Fe] abundances.  Five further objects (three in HCG\,22
and two in HCG\,90) show super-solar abundance, and one object
HCG22\_UCD001 has solar abundance.  Interestingly, the one UCD in
HCG\,22 with sub-solar abundance is HCG22\_UCD002 is the largest
UCD ($r_h\simeq 21~\rm pc$) and the only one with a spectroscopic
age estimate clearly below 12 Gyr. This is consistent with a scenario
where this object has been able to form stars over a prolonged
period of time until only a few Gyrs ago. Among all compact objects
detected, the properties of HCG22\_UCD002 differ most from those
of normal GCs.

\subsection{UCD formation}\label{UCD_formation}

The formation channels discussed for UCDs can be broken down in two
distinct concepts:

\begin{itemize}

\item UCDs are the tidally stripped remnants of extended dwarf
galaxies, hence of galactic origin and tracing the tidal disruption
of low-mass cosmological substructures.

\item UCDs were formed together with the main body of their host
galaxies' star cluster system, representing the very bright tail
of the globular cluster luminosity function \citep{mie04a,gre09}.
This encompasses both the formation of metal-rich UCDs in galaxy
mergers \citep{ash92} -- possibly as clusters of star clusters
\citep{feh02} -- and the 'primordial' channel \citep{dri04} for the
formation of metal-poor compact objects.

\end{itemize}

Regarding simple number counts, a star-cluster origin for most of
the detected UCDs is consistent with our results: the extrapolation
of the HCG\,22 GCLF to bright luminosities agrees with the number
of UCDs. Also their centrally concentrated spatial distribution
fits to this scenario.

To better distinguish between the star cluster and galactic origin
of UCDs, stellar populations are the key. (Merged) star clusters
should likely have simple and single stellar populations, corresponding
to the individual clusters out of which they merged, which likely
all formed in the same giant molecular cloud. Moreover, they should
have short star formation time-scales and therefore be enhanced in
$\alpha$ elements. In contrast, compact objects that formed within
a deeper dwarf galaxy potential are more prone to prolonged and
multiple star formation episodes, e.g. due to gas-infall. Our results
indicate a broad range of [$\alpha$/Fe] abundances for the UCDs
detected covering sub-solar to well super-solar values. One UCD
candidate in each group appears to have sub-solar alpha abundances,
such that these two are the most promising stripped dwarf galaxy
candidates. This especially holds for the HCG\,22 UCD, which in
addition to the alpha decrement has evidence for intermediate age
stellar populations and is comparably large $r_h\sim21~\rm pc$.
Supporting this scenario, the faint end of the luminosity function
of galaxies in compact groups is known to be underpopulated
\citep{hun98,men99}.

Focusing specifically on the creation of UCDs in galaxy mergers,
HCG\,90 has an interaction process happening between the three
galaxies in our targeted region, HCG\,90b, c and d, with evidences
of exchange of gas and gas disks with overlapping rotation curves
\citep{pla98,cas04}. This likely has originated the IGL component,
could have formed young star clusters, and also distributed
pre-existing objects into the IGM. Our results imply that there is
a relatively low number of UCDs in HCG\,90 in the age range of 0.5
Gyr and older, and that they are most consistent with ages of around
a Hubble time. We can thus conclude that before the very last 1-2
galaxy passages, no significant population of UCDs was formed in
HCG\,90. We are not sensitive to very young objects created after
this. Hence, for HCG\,90 our results are not sufficiently conclusive
to support or disfavour the merger scenario. HCG\,22 does not present
clear signs of ongoing interaction. However, NGC\,1199 presents a
central dust lane in its center similar to the more local case of
NGC\,5128 (CenA) of approximately $3~\rm kpc$ and about $10^7 M_{\odot}$
\citep{nie83,spa85,ver88,fin10}. This is a quite strong indication
of a gas-rich merger, old enough to have settled in a single remanent,
that likely has triggered the formation of star clusters. The
specific frequency of UCDs in HCG\,22 appears to be a few times
higher than in HCG\,90, which is hence consistent with a scenario
where the UCD population in HCG\,22 was enhanced by this merger
episode.

\vspace{0.2cm}

\noindent We conclude that the UCD population detected in each group
does in their majority not originate from relatively recent galaxy
interactions, but that rather most of the detected UCDs have been
brought into the group together with their host galaxies. While the
analysis of the objects in HCG\,90 seem to favor a formation model
where the main process happens on the group potential, the results
for HCG\,22 seem to favor models where the UCDs are formed in
processes happening in the galaxy potential, either primordial or
merger induced.

This and the existence of at least 1-2 massive and large UCDs with
intermediate ages points to the origin of UCDs being associated to
different formation processes and not a universal one.

Future investigations in this context should focus on the search
for very young UCD progenitors (few 100 Myrs) by targeting HCGs
with small crossing time scales and extending the search to bluer
colours.

\begin{acknowledgements}

We thank the referee Prof. Michael Drinkwater for his comments.
CDR would like to thank the N\'ucleo de Astrof\'{\i}sica Te\'orica
at the Universidade Cruzeiro do Sul for their hospitality, and the
support from the Deu\-tsches Zen\-trum f\"ur Luft- und Raum\-fahrt,
DLR (project number 50 OR 0602) and from the Deut\-sche
For\-schungs\-ge\-mein\-schaft, DFG (project number ZI 663/8--1)
within the Priority Program 1177 (SPP/DFG). IG is thankful for the
financial support through DFG-Projekt BO-779/32-1.

\end{acknowledgements}

\bibliographystyle{aa}

\bibliography{./ucdhcg}

%\newpage

\begin{table*}
\caption{Properties of the 16 compact stellar systems detected in
HCG\,22, ordered by $R$-band magnitude. The fifth column gives the
confidence level $R$ of the cross-correlation radial velocity
measurement, performed with the IRAF task fxcor. The last column
gives a metallicity estimate for those sources with high enough
S/N, i.e. an error of less than 0.5 dex.}
\label{table_HCG22}
\begin{center}
\begin{tabular}{l|cclrrrrc}
ID & RA  & DEC & v$_{\rm rad}$ [$\rm km~s^{-1}$]& $R$ &H$_{\beta}$ [\AA]& Mgb [\AA]& $<$Fe$>$ [\AA]&  ${\rm [Fe/H]}$ [dex]\\\hline\hline
HCG22\_UCD001 & $03^h03^m34^s46$ & $-15^o36'45''8$ & 2834 $\pm$  27 &  12.7 & 0.62 $\pm$ 0.96 & 2.97 $\pm$ 0.97 & 2.40 $\pm$ 0.70 & $-$0.3 $\pm$ 0.3\\ 
HCG22\_UCD002 & $03^h03^m36^s24$ & $-15^o37'12''8$ & 2552 $\pm$  78 &   5.1 & 3.22 $\pm$ 0.98 & 1.31 $\pm$ 1.04 & 2.18 $\pm$ 0.81 & $-$0.8 $\pm$ 0.5\\ 
HCG22\_UCD003 & $03^h03^m43^s94$ & $-15^o37'13''0$ & 2652 $\pm$  92 &   4.6 & & & & \\ 
HCG22\_UCD004 & $03^h03^m44^s62$ & $-15^o37'43''4$ & 2668 $\pm$  70 &   7.9 & 2.23 $\pm$ 1.47 & 5.91 $\pm$ 1.42 & 2.67 $\pm$ 1.03 & 0.1    $\pm$ 0.3\\ 
HCG22\_UCD005 & $03^h03^m40^s79$ & $-15^o36'57''1$ & 2665 $\pm$  51 &   8.5 & 1.44 $\pm$ 1.06 & 3.10 $\pm$ 1.03 & 1.59 $\pm$ 0.84 & $-$0.5 $\pm$ 0.4\\ 
HCG22\_UCD006 & $03^h03^m40^s74$ & $-15^o37'25''5$ & 2654 $\pm$ 118 &   4.8 & & & & \\ 
HCG22\_UCD007 & $03^h03^m34^s24$ & $-15^o37'03''2$ & 2548 $\pm$ 218 &   4.0 & & & & \\ 
HCG22\_UCD008 & $03^h03^m35^s04$ & $-15^o37'30''9$ & 2944 $\pm$  70 &   4.4 & & & & \\ 
HCG22\_UCD009 & $03^h03^m39^s55$ & $-15^o37'37''8$ & 2211 $\pm$  74 &   5.9 & & & & \\ 
HCG22\_UCD010 & $03^h03^m32^s66$ & $-15^o38'29''0$ & 2755 $\pm$ 264 &   2.9 & & & & \\ 
HCG22\_UCD011 & $03^h03^m30^s09$ & $-15^o38'09''0$ & 2766 $\pm$  95 &   4.3$^*$ & & & & \\
HCG22\_UCD012 & $03^h03^m36^s32$ & $-15^o37'15''8$ & 2549 $\pm$  35 &  10.1 & 2.65 $\pm$ 1.62 & 7.53 $\pm$ 1.50 & 4.03 $\pm$ 1.18 & 0.4    $\pm$ 0.2\\ 
HCG22\_UCD013 & $03^h03^m44^s93$ & $-15^o36'54''7$ & 2646 $\pm$ 100 &   3.2 & & & & \\ 
HCG22\_UCD014 & $03^h03^m40^s55$ & $-15^o36'18''7$ & 2905 $\pm$ 114 &   2.8 & & & & \\ 
HCG22\_UCD015 & $03^h03^m40^s18$ & $-15^o37'00''9$ & 2789 $\pm$  60 &   2.7 & & & & \\ 
HCG22\_UCD016 & $03^h03^m38^s07$ & $-15^o37'18''6$ & 2487 $\pm$ 128 &   3.1 & & & & \\ 
 \hline
\end{tabular}
\end{center}
$^*$ Cross correlation amplitude using an alternative intermediate age template (see text for details).\\

\end{table*}

%\newpage

\begin{table*}
\caption{Properties of the 5 compact stellar systems detected in
HCG\,90, ordered by $R$-band magnitude. The fifth column gives the
confidence level $R$ of the cross-correlation radial velocity
measurement, performed with the IRAF task fxcor. The last column
gives a metallicity estimate for those sources with high enough
S/N, i.e. an error of less than 0.5 dex.}
\label{table_HCG90}
\begin{center}
\begin{tabular}{l|cclrrrrc}
ID & RA   & DEC & v$_{\rm rad}$ [$\rm km~s^{-1}$]& $R$ & H$_{\beta}$ [\AA]& Mgb [\AA]& $<$Fe$>$ [\AA]& ${\rm [Fe/H]}$\\\hline\hline
HCG90\_UCD001  & $22^h01^m57^s99$ & $-31^o57'31''9$ & 2911 $\pm$  34 &  11.8 & 2.14 $\pm$ 0.62 & 3.62 $\pm$ 0.63 & 1.46 $\pm$ 0.51  & $-$0.5 $\pm$ 0.2\\ 
HCG90\_UCD002  & $22^h01^m56^s59$ & $-31^o59'25''3$ & 2567 $\pm$  41 &   9.2 & 1.80 $\pm$ 0.61 & 1.24 $\pm$ 0.65 & 2.05 $\pm$ 0.50  & $-$0.9 $\pm$ 0.3\\ 
HCG90\_UCD003 & $22^h01^m56^s42$ & $-31^o57'15''5$ & 2536 $\pm$  35 &  11.5 & 2.34 $\pm$ 0.77 & 2.85 $\pm$ 0.81 & 1.36 $\pm$ 0.63  & $-$0.6 $\pm$ 0.3\\ 
HCG90\_UCD004 & $22^h02^m01^s41$ & $-31^o55'60''0$ & 2423 $\pm$  73 &   5.5 &  & & & \\ 
HCG90\_UCD005  & $22^h02^m00^s83$ & $-31^o58'26''7$ & 3237 $\pm$  65 &   5.6 &  & & & \\ 
 \hline
\end{tabular}
\end{center}
\end{table*}

\newpage

\begin{table*}
\caption{Photometric and structural properties of HCG\,22 UCD
candidates from the HST imaging. The columns list the object ID,
foreground reddening corrected ACS Vega and ground based 
Johnson/Cousins magnitudes, half-light radius in pixels and in parsecs
calculated with DM\,=\,32.6\,mag, King concentration parameter and
the S/N of the source detection.}
\label{table_HCG22_HST}
\begin{center}
\begin{tabular}{c|ccccccccc}
ID        & F450W$_0$        & F606W$_0$        & F814W$_0$        & $B_0$ & $R_0$  & $r_{\rm h}$               & $r_{\rm h}$ & $r_{\rm t}/r_{\rm c}$ & S/N \\
	  & [mag]            & [mag]            & [mag]            & [mag] & [mag]  & [pix]                     & [pc] & & \\\hline\hline
HCG22\_UCD001 & $21.926\pm0.108$ & $20.742\pm0.083$ & $19.783\pm0.092$ & 22.56 & 21.10 & $0.63 ^{+0.013}_{-0.021}$ & 10.11 & 30 & 72.9\\ 
HCG22\_UCD002 & $21.767\pm0.150$ & $20.853\pm0.129$ & $20.127\pm0.138$ & 22.31 & 21.19 & $1.29 ^{+0.047}_{-0.046}$ & 20.71 & 15 & 56.4\\ 
HCG22\_UCD003 & -- & -- & -- & 22.46 & 21.30 & -- & -- & -- & \\ 
HCG22\_UCD004 & -- & -- & -- & 23.10 & 21.50 & -- & -- & -- & \\ 
HCG22\_UCD005 & $22.244\pm0.064$ & $21.362\pm0.051$ & $20.477\pm0.083$ & 22.91 & 21.78 & $0.11 ^{+0.055}_{-0.016}$ & 1.77 & 5 & 46.0 \\ 
HCG22\_UCD006 & -- & -- & -- & 23.17 & 21.87 & -- & -- & -- & \\ 
HCG22\_UCD007 & $22.470\pm0.078$ & $21.647\pm0.072$ & $20.797\pm0.085$ & 23.21 & 21.89 & $0.43 ^{+0.032}_{-0.025}$ & 6.90 & 5 & 42.8 \\ 
HCG22\_UCD008 & $22.601\pm0.083$ & $21.778\pm0.057$ & $20.997\pm0.075$ & 23.19 & 22.01 & $0.55 ^{+0.037}_{-0.053}$ & 8.83 & 5 & 34.6 \\ 
HCG22\_UCD009 & -- & -- & -- & 23.48 & 22.04 & -- & -- & -- &  \\ 
HCG22\_UCD010 & -- & -- & -- & 23.29 & 22.16 & -- & -- & -- &  \\ 
HCG22\_UCD011 & $23.079\pm0.074$& $22.038\pm0.071$ & $21.292\pm0.088$ & 23.35 & 22.22 & $0.31 ^{+0.063}_{-0.036}$ & 4.98 & 5 & 30.4 \\
HCG22\_UCD012 & $22.972\pm0.085$ & $21.919\pm0.052$ & $20.897\pm0.061$ & 23.83 & 22.34 & $0.44 ^{+0.032}_{-0.055}$ & 7.06 & 5 & 27.3 \\ 
HCG22\_UCD013 & -- & -- & -- & 23.77 & 22.39 & -- & -- & -- &  \\ 
HCG22\_UCD014 & $22.938\pm0.079$ & $22.018\pm0.052$ & $21.220\pm0.095$ & 23.57 & 22.47 & $0.19 ^{+0.164}_{-0.024}$ & 3.05 & 30 & 28.4 \\ 
HCG22\_UCD015 & $22.994\pm0.078$ & $22.088\pm0.056$ & $21.218\pm0.101$ & 23.80 & 22.51 & $0.40 ^{+0.204}_{-0.121}$ & 6.42 & 5 & 24.6 \\ 
HCG22\_UCD016 & $23.025\pm0.098$ & $22.118\pm0.080$ & $21.450\pm0.106$ & 23.59 & 22.54 & $0.28 ^{+0.139}_{-0.046}$ & 4.50 & 5 & 22.4 \\ 
 \hline
\end{tabular}
\end{center}
\end{table*}

\begin{table*}
\caption{Photometric and structural properties of HCG\,90 UCD
candidates from the HST imaging. The columns list the object ID,
foreground reddening corrected WFPC2 Vega and ground based
Johnson/Cousins magnitudes, half-light radius in pixels and in
parsecs calculated with DM\,=\,32.6\,mag, King concentration parameter
and the S/N of the source detection.}
\label{table_HCG90_HST}
\begin{center}
\begin{tabular}{c|ccccccccc}
ID        & F475W$_0$          & F850LP$_0$ & $B_0$ & $V_0$ & $R_0$ & $r_{\rm h}$& $r_{\rm h}$ & $r_{\rm t}/r_{\rm c}$ & S/N \\
          & [mag]              & [mag]      & [mag] & [mag] & [mag] & [pix]      & [pc]        &                       & \\
\hline
\hline
HCG90\_UCD001  & -- & -- & 22.96 & 22.00 & 21.53 & -- & -- & -- & -- \\
HCG90\_UCD002  & -- & -- & 22.93 & 22.07 & 21.62 & -- & -- & -- & -- \\
HCG90\_UCD003 & -- & -- & 22.91 & 22.10 & 21.68 & -- & -- & -- & -- \\
HCG90\_UCD004 & -- & -- & 23.05 & 22.26 & 21.87 & -- & -- & -- & -- \\
HCG90\_UCD005 & $22.978\pm0.026$ & $21.359\pm0.145$ & 23.45 & 22.61 & 22.18 & $0.39 ^{+0.016}_{-0.001}$ & 3.13 & 100 & $0.21^{+0.035}_{-0.104}$ \\
 \hline
\end{tabular}
\end{center}
\end{table*}

\Online

\begin{appendix} 

\section{High resolution colour view of our observed fields}
\begin{figure*}
\centering
\caption{Colour composite image of HCG\,22 with Keck $B$ and $R$-band
\citep{dar02} and HST/WFPC2 F450W, F606W and F814W filters. A zoom
in ($20\times20''$) showing the dust lane in NGC\,1199 can be seen
in the upper left corner.  As in Figure~\ref{images}, blue diamonds
indicate the targeted objects in our masks and red circles indicate
the HCG compact members. Field of view is $7\times5'$.}
\label{h22colour}
\end{figure*}

\begin{figure*}
\centering
\caption{Colour composite image of HCG\,90 with FORS $B$, $V$ and
$R$-band and HST/ACS F475W and F850LP filters. A zoom in
($38\farcs5\times57\farcs7$) showing the interacting pair NGC\,7174/7176
can be seen in the upper right corner.  As in Figure~\ref{images},
blue diamonds indicate the targeted objects in our masks and red
circles indicate the HCG compact members detected.  Field of view
is $7\times7'$.}
\label{h90colour}
\end{figure*}

\end{appendix}

\end{document}